\documentclass[10pt,journal,compsoc]{IEEEtran}
\usepackage{amsmath}
\usepackage{tabularx,multirow}
\usepackage{graphicx,subfigure,comment}
\usepackage[table]{xcolor}
\usepackage{url} 
\usepackage{psfrag,dsfont}
\usepackage{cite}
\usepackage{siunitx}

\usepackage[algoruled,medskip,dontprintsemicolon,linesnumbered,Algorithm]{algorithm}
\usepackage[noend]{algpseudocode}

\newcommand{\Fig}[1]{Fig.~\ref{fig:#1}}
\newcommand{\Prop}[1]{Property~\ref{prop:#1}}
\newcommand{\Ex}[1]{Example~\ref{ex:#1}}
\newcommand{\Lem}[1]{Lemma~\ref{lemma:#1}}
\newcommand{\Thm}[1]{Theorem~\ref{thm:#1}}

\newcommand{\Sec}[1]{Sec.~\ref{sec:#1}}
\newcommand{\Tab}[1]{Tab.~\ref{tab:#1}}
\newcommand{\Eq}[1]{(\ref{eq:#1})}

\newcommand{\dv}{{\bf d}}
\def\ind{{\rm 1\hspace{-0.90ex}1}}

\newcommand{\Kc}{\mathcal{K}}
\newcommand{\Hc}{\mathcal{H}}

\newcommand{\Gc}{\mathcal{G}}
\newcommand{\Vc}{\mathcal{V}}
\newcommand{\Ec}{\mathcal{E}}
\newcommand{\Qc}{\mathcal{Q}}
\newcommand{\PP}{\mathds{P}} 

\newtheorem{theorem}{Theorem}
\newtheorem{property}{Property}

\newtheorem{lemma}{Lemma}

\newtheorem{example}{Example}

\begin{document}

\title{
VNF Placement and Resource Allocation for the Support of Vertical Services in 5G Networks
} 

\author{
Satyam~Agarwal,~\IEEEmembership{Member,~IEEE,}
Francesco~Malandrino,~\IEEEmembership{Member,~IEEE,}
Carla-Fabiana~Chiasserini,~\IEEEmembership{Fellow,~IEEE,}
Swades~De,~\IEEEmembership{Senior Member,~IEEE}
\IEEEcompsocitemizethanks{\IEEEcompsocthanksitem S.~Agarwal is with IIT Guwahati, India. F.~Malandrino and C.-F.~Chiasserini are with Politecnico di Torino, Italy and CNR-IEIIT, Italy. S.~De is with IIT Delhi, India.
\IEEEcompsocthanksitem A preliminary version~\cite{noi-infocom18} of this work was presented at the IEEE INFOCOM 2018 conference.
}
}

\maketitle

\begin{abstract}
One of the main goals of 5G networks is to support the  technological and business needs of various industries (the so-called verticals), which wish to offer to their customers a wide range of services characterized by diverse performance requirements. In this context, a critical challenge lies in mapping in an automated manner the requirements of verticals into decisions concerning the network infrastructure, including VNF placement, resource assignment, and traffic routing. 
In this paper, we seek to make such decisions {\em jointly}, accounting for their mutual interaction, and efficiently. To this end, we formulate a queuing-based model and use it at the network orchestrator to optimally match the vertical's requirements to the available system resources. We then propose a fast and efficient solution strategy, called MaxZ, which allows us to reduce the solution complexity. Our performance evaluation, carried out accounting for multiple scenarios representative of real-world services, shows that MaxZ performs substantially better than state-of-the-art alternatives and consistently close to the optimum.

\end{abstract}

\section{Introduction}
\label{sec:intro}

5G networks are envisioned to 
provide the computational, memory, and storage resources needed to run
multiple third parties (referred to as vertical industries or {\em verticals}) with diverse communication and computation needs. Verticals provide network operators with the specification of the services they want to provide, e.g., the virtual network functions (VNFs) they want to use to process their data and the associated quality of service. 

Mobile network operators are in charge of
mapping the requirements of the verticals into infrastructure management decisions. This task is part of the network {\em orchestration}, and includes making decisions concerning (i) the {\em placement} of the VNFs needed by the verticals across the infrastructure; (ii) the {\em assignment} of CPU, memory and storage resources to the VNFs; (iii) the {\em routing} of data across network nodes.

These decisions interact with each other in ways that are complex and often counterintuitive. In this paper, we focus on the allocation of computational and network resources, and make such decisions jointly, accounting for (i) the requirements of each VNF and vertical; (ii) the capabilities of the network operator's infrastructure; (iii) the capacity and latency of the links between network nodes.
A key aspect of our work, often disregarded by previous literature
on 5G and VNF placement,
is that our approach allows {\em flexible allocation} of the computational capabilities of each host among the VNFs it runs.

We identify  queuing theory as the best tool to model 5G networks, owing to the nature of their traffic and the processing such a traffic needs. Indeed:
\begin{itemize}
    \item much of 5G traffic, especially that coming from Internet-of-things (IoT) and machine-type communication (MTC) applications, will consist of REST-ful, atomic (in principle) requests, as opposed to long-standing connections~\cite{3gppmtc};
    \item such requests will traverse one or more processing stages, as implemented in the emerging multi-access edge computing (MEC) implementation Amazon Greengrass~\cite{greengrass}, and can trigger additional requests in the process;
    \item the time it takes to process each request depends on the capabilities of the computational entity serving it~\cite{greengrass}.
\end{itemize}
Requests and processing stages naturally map onto clients and queues they have to traverse. Furthermore, the fact that queues can be assigned different service rates aptly models our flexible allocation of computational resources.

We take {\em service delay} as our main key performance indicator (KPI), and we formulate an optimization problem that minimizes the maximum ratio between actual and maximum allowed end-to-end latency, across all services.
Furthermore, and without loss of generality, we focus on CPU as the resource to assign to VNFs.
In light of the complexity of the problem, we then propose an efficient solution strategy,
closely matching the optimum:
based on (i) {\em decoupling} the VNF placement and CPU assignment decisions, while keeping track of their interdependence, and (ii) {\em sequentially} making such decisions for each VNF. Traffic routing decisions are simply derived once all placement and assignment decisions are made.
Although made in a decoupled and sequential fashion, our decisions are joint as their mutual impact is properly accounted for, e.g., we consider how deploying a new VNF on a host impacts the possible CPU assignments therein.

Our main contributions can be summarized as follows:
\begin{itemize}
    \item our model accounts for the main resources of 5G networks, namely, hosts and links;
    \item we model the diverse requirements of different VNFs, and allow them to be composed in arbitrarily complex graphs, as mandated by~\cite[Sec.~6.5]{etsimano}, instead of simpler chains or directed acyclic graphs (DAGs);
    \item unlike existing work, we allow {\em flexible} allocation of CPU to VNFs, and model the resulting impact on service times;
    \item we propose a solution strategy, called MaxZ, that is able to efficiently and effectively make VNF placement and CPU allocation decisions, and show how it consistently performs very close to the optimum across a variety of traffic requirements;
    \item focusing on the special case of fully-load conditions, we state and prove several properties of the optimal CPU allocation decisions, and use them to further speed up the decision process.
\end{itemize}

The remainder of the paper is organized as follows. \Sec{relwork} reviews related work, highlighting the novelty of our contribution.
\Sec{mano} positions our work within the context of the ETSI management and orchestration (MANO) framework. 
\Sec{model} describes the system model, while \Sec{problem} introduces the problem formulation and analyzes its complexity.  \Sec{solution} presents our solution concept, while \Sec{fully} describes how we deal with the special case of full-load conditions. \Sec{manyinstances} addresses scenarios with multiple VNF instances. Finally, \Sec{results} presents performance evaluation results, while \Sec{conclusion} concludes the paper.

\section{Related work}
\label{sec:relwork}

\textbf{Network slicing and orchestration.}
A first body of works concerns the network slicing paradigm and its role within 5G. Several works, including~\cite{slicing-survey,slicing1,slicing2}, focus on the architecture of 5G networks based on network slicing, pointing out their opportunities and challenges. Other works, e.g.,~\cite{slicingp2,slicingalgos}, address decision-making in 5G networks and the associated challenges, including computational complexity. Finally, orchestration, including the decision-making involved entities and the arising security concerns have been tackled in, e.g., \cite{orch-arch} and \cite{orch-sec}, respectively.

\textbf{Network-centric optimization.}
Many works, including~\cite{AHirwe16,TKuo16,ABaumgartner15, FJemaa16,BAddis15}, tackle the problems of VNF placement and routing from a network-centric viewpoint, i.e., they aim at minimizing the load of network resources. In particular,~\cite{AHirwe16} seeks to balance the load on links and servers, while~\cite{TKuo16} studies how to optimize routing to minimize network utilization. The above approaches formulate mixed-integer linear programming (MILP) problems and propose heuristic strategies to solve them. \cite{ABaumgartner15}, \cite{FJemaa16}, and~\cite{BAddis15} formulate ILP problems, respectively aiming at minimizing the cost of used links and network nodes, minimizing resource utilization subject to QoS requirements, and minimizing bitrate variations through the VNF graph.

\textbf{Service provider's perspective.}
Several recent works take the viewpoint of a service provider, supporting multiple services that require different, yet overlapping, sets of VNFs, and seek to maximize its revenue. The works~\cite{AMarotta16,NKhoury16} aim at minimizing the energy consumption resulting from VNF placement decisions. \cite{MMechtri16,LGu16}~study how to place VNFs between network-based and cloud servers so as to minimize the cost, and \cite{JCao17}~studies how to design the VNF graphs themselves, in order to adapt to the network topology.

\textbf{User-centric perspective.}
Closer to our own approach, several works take a user-centric perspective, aiming at optimizing the user experience. \cite{infocom15_optimal,martini2015latency}~study the VNF placement problem, accounting for the computational capabilities of hosts as well as network delays. In~\cite{DBhamare17}, the authors consider inter-cloud latencies and VNF response times, and solve the resulting ILP through an affinity-based heuristic.

\textbf{Virtual EPC.}
The Evolved Packet Core (EPC) is a prime example of a service that can be provided through software defined networking and network function virtualization (SDN/NFV). Interestingly, different works use different VNF graphs to implement EPC, e.g., splitting user- and control-plane entities~\cite{epc_split,globecom15_bearer,vepc_controller_placement} or joining together the packet and service gateways (PGW and SGW)~\cite{comsnets17_placement,tvt_mme}. Our model and algorithms work with any VNF graph, which allows us to model any real-world service, including all implementations of vEPC.

\subsection{Novelty}

The closest works to ours, in terms of approach and/or methodology, are~\cite{infocom15_optimal}, \cite{martini2015latency}, \cite{DBhamare17}, and~\cite{comsnets17_placement}.

In particular,~\cite{infocom15_optimal}, \cite{martini2015latency}, and~\cite{comsnets17_placement} model the assignment of VNFs to servers as a generalized assignment problem, a resource-constrained shortest path problem and a MILP problem, respectively. This implies that either a server has enough spare CPU capacity to offer a VNF, or it does not. Our queuing model, instead, is the first to account for the flexible allocation of CPU to the VNFs running on a host, e.g., the fact that VNFs will work faster if placed at a scarcely-utilized server. Furthermore, \cite{infocom15_optimal} and~\cite{comsnets17_placement} have as objective the minimization of costs and server utilization, respectively. Our objective, instead, is to minimize the delay incurred by requests of different classes, which changes the solution strategy that can be adopted.
The work~\cite{martini2015latency} aims at solving essentially the same problem as ours, albeit in the specific scenario where all traffic flows through a deterministic sequence of VNFs, i.e., VNF graphs are chains.

The queuing model used in~\cite{DBhamare17} is similar (in principle) to ours; however, \cite{DBhamare17} does not address overlaps between  VNF graphs and only considers DAGs,
i.e., requests cannot visit the same VNF more than once.
Furthermore, in both~\cite{martini2015latency} and~\cite{DBhamare17} no CPU allocation decisions are made, and the objective is to minimize a global metric, ignoring the different requirements of different service classes. Finally, the affinity-based placement heuristic proposed in~\cite{DBhamare17} neglects the inter-host latencies, and this, as confirmed by our numerical results in \Sec{results}, can yield suboptimal performance.

Finally, it is worth mentioning that a preliminary version of this paper appeared in~\cite{noi-infocom18}. While sharing the same basic solution concept, this version includes a substantial amount of new and revised material, including a discussion on how our work fits in the 5G~MANO framework (\Sec{mano}), an extended discussion of full-load conditions (\Sec{fully}), and new results for large-scale scenarios.

\section{Our work and the ETSI MANO framework}
\label{sec:mano}

ETSI has standardized \cite{etsimano} the management and orchestration (MANO) framework, including a set of functional blocks and the reference points, i.e., the interfaces between functional blocks (akin to a REST API) that they use to communicate. Its high-level purpose is to translate business-facing KPIs chosen by the vertical (e.g., the type of processing needed and the associated end-to-end delay) into resource-facing decisions such as virtual resource instantiation, VNF placement, and traffic routing. In this section, we first present a brief overview of the ETSI MANO framework; then, in \Sec{mano-r-us}, we focus on the NFV orchestrator and detail the decisions it has to make and the input data at its disposal.


\begin{figure*}
\centering
\includegraphics[width=1.5\columnwidth]{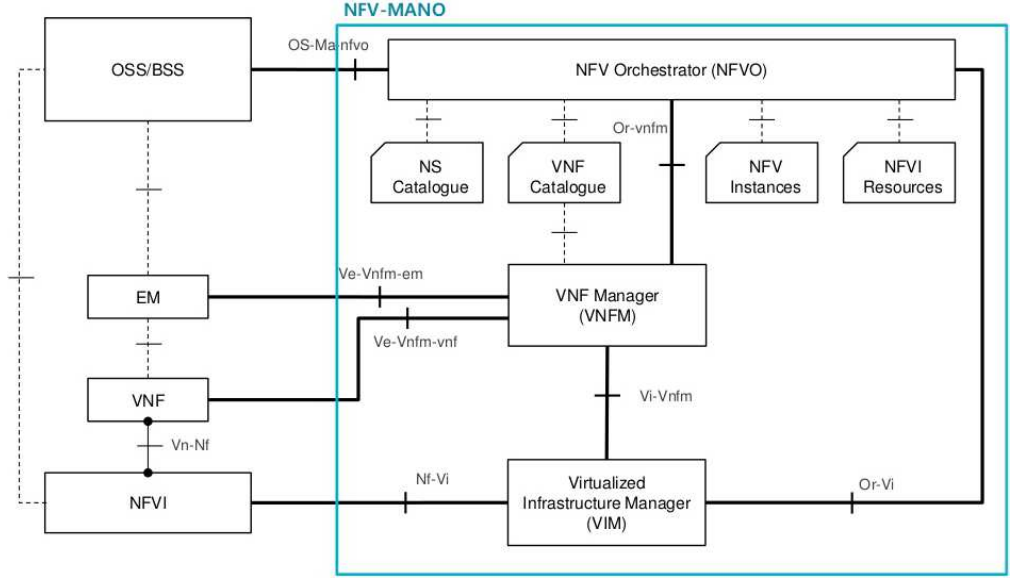}
\caption{The NFV-MANO architectural framework. Source: \cite{etsimano}
\label{fig:mano}
}
\end{figure*}

\Fig{mano} presents the functions composing the MANO framework (within the blue area) as well as the functions outside the framework they interact with. Operation and business support (OSS/BSS) service block, which represent the interface between verticals and mobile operators. High-level, end-to-end requirements and KPIs are conveyed, through the Os-Ma-nfvo reference point, to the NFV orchestrator (NFVO). The NFVO is in charge of deciding the number and type of VNFs to instantiate as well as the capacity of virtual links (VLs) connecting them.

Such decisions are conveyed, via the Or-vnfm interface, to the VNF manager (VNFM) function, which is in charge of actually instantiating the required VNFs. The VNFM requests from the virtual infrastructure manager (VIM) any resource, e.g., virtual machine (VM) or VL needed by the VNFs themselves.
The VNFM also interacts with the element management (EM) function,
a non-MANO entity that is in charge of Fault, Configuration, Accounting, Performance and Security (FCAPS) management for the functional part of the VNFs, i.e., for the actual tasks they perform.

Finally, the VIM interacts with the NFV infrastructure (NFVI), which includes the hardware (e.g., physical servers, network equipment, etc.) and software (e.g., hypervisors) running the VNFs.

\subsection{The NFVO: input, output, and decisions}
\label{sec:mano-r-us}

As its name suggests, the main entity in charge of orchestration decisions is the NFV orchestrator (NFVO), which belongs to the MANO framework depicted in \Fig{mano}. In the following, we provide more details on the decisions the NFVO has to make and the information it can rely upon, which correspond (respectively) to the output and input of our algorithms.

The NFVO receives from the OSS/BSS a data structure called {\em network service descriptor} (NSD), defined in~\cite[Sec.~6.2.1]{etsimano}. NSDs include a graph-like description of the processing each service requires, e.g., the VNFs that the traffic has to traverse, in the form of a {\em VNF Forwarding Graph (VNFFG)} descriptor~\cite[Sec.~6.5.1]{etsimano}. They contain {\em deployment flavor} information, including the delay requirements associated with every service~\cite[Sec.~6.2.1.3]{etsimano}. Additionally, from the virtual infrastructure manager (VIM), the NFVO fetches information on the state and availability of network infrastructure, including VMs able to run the VNFs and the links connecting them.

With such information, the NFVO can make what ETSI calls {\em lifecycle management} decisions~\cite[Sec.~7.2]{etsimano} about the VNFs composing each network slice, i.e., how many instances of these VNFs to instantiate, where to host them, and how much resources to assign to each of them. Such decisions will correspond to decision variables in our system model, as detailed next. 

\section{System model}
\label{sec:model}

We model VNFs as M/M/1 {\em queues}, belonging to set~$\Qc$, whose customers correspond to service requests. The {\em class} of each customer corresponds to the service with which each request is associated; we denote the set of such classes by~$\Kc$. The {\em service rate}~$\mu(q)$ of each queue~$q$ reflects the
amount of CPU (expressed in, e.g., ticks or microseconds of CPU-time)
each VNF is assigned to. Thus, $\mu(q)$~influences the time taken to process one service request.
Notice that~$\mu(q)$ does not depend on the class~$k$; that is, CPU is assigned on a per-VNF rather than per-class basis. This models those scenarios where the same VNF instance can serve requests belonging to multiple services.

{\em Arrival rates} at queue $q\in\Qc$ are denoted by~$\lambda_k(q)$. Note that these values are class-specific, and reflect the  amount of traffic of different services. Class-specific {\em transfer probabilities}~$\PP(q_2|q_1,k)$ indicate the probability that a service request of class~$k$  enters  VNF~$q_2$ after being served by VNF~$q_1$. Furthermore,~$\PP(q|\circ,k)$ indicates the probability that a request of class~$k$ starts its processing at VNF~$q$.

Physical, or more commonly virtual, {\em hosts} are represented by set~$\Hc$. Each host~$h$ has a finite {\em CPU capacity}~$\kappa_h$. Host-specific $\kappa_h$~values account for both  different capabilities and different hosts, and the fact that some hosts may be assigned a low-power CPU state~\cite{cstates} for energy-saving purposes. This implies that energy constraints can be accounted for by properly setting the values of the $\kappa_h$ parameters.

Going from host~$h\in\Hc$ to host~$l\in\Hc$ entails a deterministic {\em network latency}~$\delta(h,l)$, which depends on the (virtual) link between the two hosts. Furthermore, the link between hosts~$h$ and~$l$ has a finite capacity~$C(h,l)$.
The delay and capacity parameters, i.e.,~$\delta(h,l)$ and~$C(h,l)$, are able to describe in a consistent way two different cases, namely:
\begin{itemize}
    \item there is a direct, physical link between hosts~$h$ and~$l$;
    \item $h$ and~$l$ are connected through multiple physical links, which are abstracted as one virtual link.
\end{itemize}
In the latter case, the capacity of the virtual link corresponds to that of the physical link with lowest capacity, while the delay is the sum of individual delays. Such information is part of the input data to our problem.

The notation used is summarized in \Tab{notation}, and exemplified below.

\begin{figure}[t]
\footnotesize
\psfrag{l1}[r][t]{$\lambda_\text{video}$}
\psfrag{l2}[r][c]{$\lambda_\text{game}$}
\psfrag{l3}[r][b]{$\lambda_\text{veh}$}
\psfrag{q1}[m][c]{\textsf{firewall}}
\psfrag{q2}[m][c]{\textsf{   transcoder}}
\psfrag{q3}[m][c]{\textsf{    game server}}
\psfrag{q4}[m][c]{\textsf{     coll. detector}}
\psfrag{q5}[m][c]{\textsf{billing}}
\psfrag{q6}[m][c]{\textsf{DPI}}
\psfrag{c1}[m][c]{\textsf{$h$}}
\psfrag{c2}[m][c]{\textsf{$l$}}
\psfrag{c3}[m][c]{\textsf{$m$}}
\psfrag{d1}[t][c]{\textsf{$\delta(h,l)$}}
\psfrag{d2}[t][c]{\textsf{$\delta(l,m)$}}
\psfrag{e1}[t][c]{\textsf{$C(h,l)$}}
\psfrag{e2}[t][c]{\textsf{$C(l,m)$}}

\centering
\includegraphics[width=1\columnwidth]{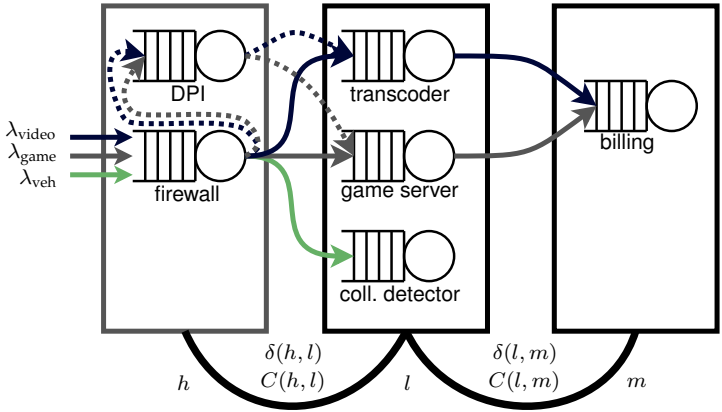}
\caption{\Ex{simple}: three service graphs, and six VNFs, corresponding to the six queues, placed across three hosts.
Dashed and dotted lines represent the different paths that service requests can take.
\label{fig:example}
} 
\end{figure}

\noindent\rule{1\columnwidth}{0.5mm}
\begin{example}
    \label{ex:simple}
    Assume that the network has to support three services: video streaming, gaming, and vehicle collision detection.
Then the set of service classes is~$\Kc=\{\text{video},\text{game},\text{veh}\}$.
For the sake of clarity, let us associate to each service the following, highly simplified, VNF graphs:
    \begin{itemize}
        \item {\em video streaming}: firewall -- transcoder -- billing;
        \item {\em gaming}: firewall -- game server -- billing;
        \item {\em vehicle collision detection}: firewall -- collision detector.
    \end{itemize}

    Suspicious-looking packets belonging to the video streaming and gaming services can further be routed through a deep packet inspection (DPI) VNF. Hence, $\Qc$=$\{$\text{firewall}, \text{transcoder}, \text{billing}, \text{game server}, \text{collision detector}, \text{DPI}$\}$. 

    There are three hosts~$\Hc=\{h,l,m\}$, connected to each other through links characterized by a latency $\delta$ and a  link capacity $C$.
    \Fig{example} illustrates the above quantities and shows a possible VNF placement across the three hosts.
    Routing can be  deterministic, e.g., $\PP(\text{billing}| \text{transcoder},\text{video})$=$1$, or it can be  probabilistic, e.g.,  $\PP(\text{DPI} | \text{firewall},\text{gaming})$=$0.1$ and $\PP(\text{game server} | \text{firewall},\text{gaming})$=$0.9$.
\end{example}
\noindent\rule{1\columnwidth}{0.5mm}

\begin{table}
\caption{Notation
\label{tab:notation}
}
\begin{tabularx}{\columnwidth}{|l|l|X|}
\hline
$A(h,q)$ & Binary variable & Whether to deploy VNF~$q$ at host~$h$ \\
\hline
$C(h,l)$ & Parameter & Capacity of the link between hosts~$h$ and~$l$ \\
\hline
$D_k$ & Aux. var. & Total delay incurred by requests of class~$k$ \\
\hline
$D_k^\text{QoS}$ & Parameter & Target delay for requests of class~$k$ \\
\hline
$h\in\Hc$ & Set & Physical hosts \\
\hline
$k\in\Kc$ & Set & Services (traffic classes) \\
\hline
$q\in\Qc$ & Set & VNFs (queues) \\
\hline
$\PP(q_1|q_2,k)$ & Parameter & Probability that requests of class~$k$ visit VNF~$q_2$ (immediately) after VNF~$q_1$ \\
\hline
$R_k(q)$ & Aux. var. & Processing time for requests of class~$k$ at VNF~$q$ \\
\hline
$\gamma_k(q)$ & Aux. var. & Number of times requests of class~$k$ visit VNF~$q$ \\
\hline
$\delta(h,l)$ & Parameter & Delay associated with the link between hosts~$h$ and~$l$ \\
\hline
$\Lambda_k(q)$ & Aux. var. & Rate at which requests of any class arrive at VNF~$q$ \\
\hline
$\lambda_k(q)$ & Aux. var. & Rate at which {\em new} requests of class~$k$ arrive at VNF~$q$ \\
\hline
$\hat{\lambda}_k(q)$ & Aux. var. & Total rate at which requests of class~$k$ arrive at VNF~$q$ \\
\hline
$\kappa_h$ & Parameter & Computational capacity of host~$h$ \\
\hline
$\mu(q)$ & Real variable & Computational capacity to assign to VNF~$q$ \\
\hline
\end{tabularx}
\end{table}

\section{Problem formulation and complexity}
\label{sec:problem}

When allocating the resources necessary to run each service, the NFVO has to make three main decisions:
\begin{itemize}
    \item VNF placement, i.e., which physical hosts have to run the required VNFs;
    \item CPU assignment, i.e., how the computational capabilities of each host have to be shared between VNF;
    \item how traffic shall be steered between VNFs.
\end{itemize}
In this section, we describe how such decisions and the constraints they are subject to can be described through our system model.
We take service delay as our main performance metric, and we  formulate the problem for scenarios where exactly one instance of each VNF is to be deployed in the network. The general case where multiple instances of the same VNF can be deployed is then discussed in \Sec{manyinstances}.

{\bf Decisions and decision variables.} 
We have two main decision variables: a binary variable~$A(h,q)\in\{0,1\}$ represents whether VNF~$q\in\Qc$ is deployed at host~$h\in\Hc$, and a real variable~$\mu(q)$ expresses the amount of CPU assigned to VNF~$q\in\Qc$. Notice how $\mu(q)$ maps onto the service rate of the corresponding queue.

{\bf System constraints.}
A first, basic constraint that allocation decisions must meet is that the computational capacity of hosts is not exceeded, i.e.,
\begin{equation}
\label{eq:am}
\sum_{q\in\Qc} A(h,q) \mu(q) \leq \kappa_h, \quad\forall h\in\Hc.
\end{equation}
As mentioned above, we present our model in the case where there is exactly one instance of each VNF deployed in the system. This translates into:
\begin{equation}
\label{eq:onehost}
\sum_{h\in\Hc}A(h,q)= 1,\quad\forall q\in\Qc.
\end{equation}

{\bf Arrival rates and system stability.}
In addition not to overload hosts, allocation decisions must ensure that individual VNFs are assigned enough computational capacity to cope with their load, i.e., that the system is {\em stable}.
Recall that input parameters~$\lambda_k(q)$ express the rate at which {\em new} requests of service class~$k$ arrive at queue~$q\in\Qc$. We can then define an auxiliary variable~$\hat{\lambda}_k(q)$, expressing the {\em total} rate of requests of class~$k$ that enter queue~$q$, either from outside the system or from other queues. For any $k \in \Kc$, we have:
\begin{equation}
\label{eq:rates}
\hat{\lambda}_k(q) {=} \sum_{q\in\mathcal{Q}} \lambda_{k,q} + \sum_{p\in\Qc}\PP(q|p,k)\hat{\lambda}_k(p).
\end{equation}
We can then define another auxiliary variable~$\Lambda(q)$, expressing the total arrival rate of requests of any class entering queue~$q$:
\begin{equation}
\nonumber
\Lambda(q)=\sum_{k\in\Kc}\hat{\lambda}_k(q).
\end{equation}
Using $\Lambda(q)$, we can impose {\em system stability}, requesting that, for each queue, the arrival rate does not exceed the service rate:
\begin{equation}
\label{eq:mu-gt-lambda}
\Lambda(q) < \mu(q), \quad\forall q \in \Qc.
\end{equation}
In other words, each VNF should receive {\em at least} enough CPU  to deal with the incoming traffic. If additional CPU  is available at the host, it will be exploited to further speed up the processing of requests.

{\bf Latency.}
The previous constraints ensure that individual VNFs are stable, i.e., they process incoming requests in a finite time. We can now widen our focus, and study how the processing times of different VNFs and the network times combine to form our main metric of interest, i.e., the {\em delay} each request is subject to.

The processing time, i.e., the time it takes for a request of service~$k$ to traverse VNF~$q$ is represented by an auxiliary variable~$R_k(q)$. For FCFS (first come, first serve) and PS (processor sharing) queuing disciplines, we have:
\begin{equation}
\label{eq:resptime}
R_k(q) = \frac{1}{\mu(q) - \Lambda(q)},\quad\forall q\in\Qc
\end{equation}
Note that the right-hand side of \Eq{resptime} does not depend on  class~$k$; intuitively, this is because the queuing disciplines we consider are unaware of service classes. The response times for other queuing disciplines, including those accounting for priority levels and/or preemption, cannot be expressed in closed form. It is also worth stressing that present-day implementations of multi-access edge computing (MEC)~\cite{greengrass} are based on FIFO discipline, and do not support preemption.

To compute the network latency that requests incur  when transiting between hosts, we first need the expected number of times, $\gamma_k(q)$, that a request of class~$k$ visits VNF~$q\in\Qc$, i.e.,
\begin{equation}
\label{eq:gamma-k}
\gamma_k(q) = \PP(q|\circ,k) + \sum_{p \in \Qc \setminus \{q\} } \PP(q|p,k) \gamma_k(p).
\end{equation}
In the right-hand side  of \Eq{gamma-k}, the first term is the probability that requests start their processing at queue~$q$, and the second is the probability that requests arrive there from another queue~$p$. Note that~$\gamma_k(q)$ is not an auxiliary variable, but a quantity that can be computed offline given the transfer probabilities~$\PP$.
Using~$\gamma_k(q)$, the
expected
network latency incurred by requests of service class~$k$ is:
\begin{equation}
\label{eq:prop-delay}
\sum_{q,r\in\Qc}\gamma_k(q)\PP(r|q,k)\sum_{h,l\in\Hc}\delta(h,l) A(h,q)A(l,r).
\end{equation}
We can read \Eq{prop-delay} from left to right, as follows. Given a service request of class~$k$, it will be processed by VNF~$q$  for  
$\gamma_k(q)$ number of times. Every time, it will move to  VNF~$r$ with probability~$\PP(r|q,k)$. So doing, it will  incur latency~$\delta(h,l)$ if $q$ and $r$ are deployed at hosts $h$ and $l$, respectively (i.e., if $A(h,q)=1$ and $A(l,r)=1$).

The average total delay of requests of the generic service class~$k$ is therefore given by:
\begin{multline}
\label{eq:dk1}
D_k=\sum_{q\in\Qc}\gamma_k(q)R_k(q)+\\
\sum_{q,r\in\Qc, q\neq r}\gamma_k(q)\PP(r|q,k)
\sum_{h,l\in\Hc}A(h,q)A(l,r)\delta(h,l).
\end{multline}

{\bf Link capacity.}
Given the finite link capacity $C(h,l)$, which limits the number of requests that move from any VNF at host $h$ to any VNF at host~$l$,we have:
\begin{equation}
\label{eq:bandwidth}
\sum_{k\in\Kc}\sum_{q,r\in\Qc}\hat{\lambda}_k(q)\PP(r|q,k)A(q,h)A(r,l){\leq}C(h,l).
\end{equation}
Constraint \Eq{bandwidth} contains a summation over all classes~$k$ and all VNFs~$q,r\in\Qc$, such that $q$~is deployed at~$h$ and $r$~is deployed at~$l$, as expressed by the $A$-variables. For each of such pair of VNFs, $\hat{\lambda}_k(q)$~is the rate of the requests of class~$k$ that arrive at~$q$. Multiplying it by~$\PP(r|q,k)$, we get the rate at which requests move from VNF~$q$ to VNF~$r$, hence from host~$h$ to host~$l$.

{\bf Objective.}
$D_k$ defined above represents the average delay incurred by requests of class~$k$. In our objective function, we have to combine these values in a way that reflects the differences between such classes, most notably, their different QoS limits. Thus, we consider for each class~$k$ the {\em ratio} of the actual delay~$D_k$ to the limit delay~$D_k^\text{QoS}$, and seek to minimize the maximum of such ratios:
\begin{equation}
\label{eq:obj}
\min_{A,\mu}\max_{k\in\Kc}\frac{D_k}{D_k^\text{QoS}}.
\end{equation}
Importantly, the above objective function not only ensures fairness among service classes while accounting for their limit delay, but 
it also guarantees that the optimal solution will match {\em all} QoS limits if possible.
More formally:
\begin{property}
\label{prop:resilient}
If there is a non-empty set of solutions that meet constraints \Eq{am}--\Eq{bandwidth} and honor the services QoS limits, then the  optimal solution to \Eq{obj} falls in such a set.
\end{property}
\begin{IEEEproof}
We prove the property by contradiction, and assume that there is a feasible solution such that~$D^\prime_k\leq D_k^\text{QoS}$ for all service classes, but that the optimal solution has~$D_{\hat{k}}^{\star}>D_{\hat{k}}^\text{QoS}$ for at least one class~$\hat{k}\in\Kc$.

In this case, the optimal value of the objective \Eq{obj} would be at least~$\frac{D_{\hat{k}}^{\star}}{D_{\hat{k}}^\text{QoS}}>1$. However, we know by hypothesis that there is a feasible solution where~$D_k\leq D_k^\text{QoS}$ for all classes, which would result in an objective function value of~$\min_{k\in\Kc}\frac{D^\prime_k}{D_k^\text{QoS}}\leq 1$. It follows that the solution we assumed to be optimal cannot be so.
\end{IEEEproof}

Furthermore, when no solution meeting all QoS limits exists, the solution optimizing \Eq{obj} will minimize the damage by keeping all delays as close as possible to their limit values.

\subsection{Problem complexity}
The VNF placement/CPU assignment problem is akin to max-flow problem; however, it has a much higher complexity due to the following: (i) binary variables control whether edges and nodes are activated, and (ii) the cost associated with edges changes according to the values of said variables. More formally, the problem of maximizing
\Eq{obj} subject to constraints \Eq{am}--\Eq{bandwidth} includes both binary~($A(h,q)$) and continuous ($\mu(q)$) variables. More importantly, constraints \Eq{am} and \Eq{bandwidth}, as well as  objective \Eq{obj} (see also \Eq{dk1}), are nonlinear and non-convex, as both include products between different decision variables.

Below we  prove that such a problem is NP-hard,
through a reduction from the generalized assignment problem (GAP).
\begin{theorem}
\label{thm:nphard}
The problem of joint VNF placement and CPU assignment is NP-hard.
\end{theorem}
\begin{IEEEproof}
It is possible to reduce the GAP, which is NP-hard~\cite{cattrysse1992survey}, to ours. In other words, we show that (i) for each instance of the GAP problem, there is a corresponding instance of our VNF placement problem, and (ii) that the translation between them can be done in polynomial time.

\noindent{\bf GAP instance.}
The GAP instance includes {\em items}~$i_1,\dots,i_N$ and {\em bins}~$b_1,\dots,b_M$. Each bin~$b$ has a budget (size)~$s_b$; placing item~$i$ at bin~$b$ consumes a budget (weight)~$w_{bi}$ and yields a cost~$p_{bi}$. The decision variables are binary flags~$x_{bi}$ stating whether item~$i$ shall be assigned to bin~$b$; also, each item shall be assigned to exactly one bin. The objective is to minimize the cost.

\noindent{\bf Reduction.}
In our problem, items and bins correspond to VNFs and hosts respectively, and the decision variables~$x_{bi}$ correspond to VNF placement decisions~$A(i,b)$. The capacity of each host is equal to the size~$s_b$ of the corresponding bin. Furthermore, we must ensure that:
\begin{itemize}
    \item the weight~$w_{bi}$ of item~$i$ when placed at bin~$b$ corresponds to the quantity of CPU assigned to VNF $i$, i.e.,~$w_{bi}=\mu^b(i)$;
    \item the cost~$p_{bi}$ coming from placing item~$i$ in bin~$b$ corresponds to the opposite\footnote{So that minimizing the cost is the same as minimizing the service time.} of the processing time at VNF~$i$, i.e.,~$w_{bi}=-\frac{1}{\mu^b(i)-\Lambda(i)}$, or equivalently, with a linear equation,~$\Lambda(i)-\mu^b(i)=\frac{1}{p_{bi}}$.
\end{itemize}
Finally, we set all inter-host delays to zero.

\noindent{\bf Complexity of the reduction.}
Performing the reduction described above only requires to solve a linear system of equations in the~$\mu^b(i)$ and~$\Lambda(i)$ variables, which can be performed in polynomial (indeed, cubic) time~\cite{barrett1994templates}. We have therefore presented a polynomial-time reduction of any instance of the GAP problem to our problem. It follows that our problem is NP-hard, q.e.d.
\end{IEEEproof}
It is interesting to notice how, in the proof of \Thm{nphard}, we obtain a {\em simplified} version of our problem, with non-flexible CPU assignment (if VNF~$i$ is placed at host~$b$ it gets exactly~$\mu^b(i)$ CPU) and no network delay. This suggests that our problem is indeed more complex than GAP.

The NP-hardness of the problem
rules out not only the possibility to directly optimize the problem through a solver, but also commonplace solution strategies based on {\em relaxation}, i.e., allowing binary variables to take values anywhere in~$[0,1]$. Even if we relaxed the $A(h,q)$~variables, we would still be faced with a non-convex formulation, for which no algorithm is guaranteed to find a global optimum.

One approach to overcome such an issue could be simplifying the model, e.g., by assuming that any host has sufficient computing capability to run simultaneously all VNFs, therefore dispensing with the $A(h,q)$~variables. However, by doing so we would detach ourselves from real-world 5G systems, thus jeopardizing the validity of the conclusions we draw from our analysis. We instead opt to keep the model unchanged and
present an efficient, {\em decoupled} solution strategy, leveraging on sequential decision making.

\section{Solution strategy}
\label{sec:solution}

Our solution strategy is based on decoupling the problems of VNF placement and CPU allocation, and then sequentially
-- and yet jointly, i.e., accounting for their mutual impact --
making these decisions. We begin by presenting our VNF placement heuristic, called {\em MaxZ}, in \Sec{sol-maxz}, and then discuss CPU allocation in \Sec{sol-mu}.

\subsection{The MaxZ placement heuristic}
\label{sec:sol-maxz}

As mentioned earlier, the two main sources of problem complexity are binary variables and non-convex functions in both  objective \Eq{obj} and  constraints \Eq{am} and \Eq{bandwidth}. In order to solve the VNF placement problem, our heuristic walks around these issues by:
\begin{enumerate}
    \item formulating a convex version of the problem;
    \item solving it through an off-the-shelf solver;
    \item computing, for each VNF~$q$ and host~$h$, a {\em score}~$Z(h,q)$, expressing how confident we feel about placing~$q$ in~$h$;
    \item considering the maximum score~$Z(h^\star,q^\star)$ and placing VNF~$q^\star$ at host~$h^\star$;
    \item repeating steps~2--4 until all VNFs are placed.
\end{enumerate}
The name of the heuristic comes from step~4, where we seek for the highest score~$Z$.

\subsubsection{Steps 1--2: convex formulation}
\label{sec:maxz-convex}

To make the problem formulation in \Sec{problem} convex, first we need to get rid of binary variables; specifically, we replace the binary variables~$A(h,q)\in\{0,1\}$ with continuous variables~$\tilde{A}(h,q)\in[0,1]$.

We also need to remove the products between $\tilde{A}$-variables (e.g., in \Eq{prop-delay}, \Eq{dk1}, and \Eq{bandwidth}), by replacing them with a new variable. To this end, for each pair of VNFs~$q$ and~$r$ and hosts~$h$ and~$l$, we introduce a new variable~$\Phi(h,l,q,r)\in[0,1]$, and impose that:

\begin{equation}
\label{eq:phi-1}
\Phi(h,l,q,r)\leq \tilde{A}(h,q),\quad\forall h,l\in\Hc,q,r\in\Qc;
\end{equation}
\begin{equation}
\label{eq:phi-2}
\Phi(h,l,q,r)\leq \tilde{A}(l,r),\quad\forall h,l\in\Hc,q,r\in\Qc;
\end{equation}
\begin{equation}
\label{eq:phi-3}
\Phi(h,l,q,r){\geq} \tilde{A}(h,q){+}\tilde{A}(l,r){-}1,\quad\forall h,l\in\Hc,q,r\in\Qc.
\end{equation}
The intuition behind constraints \Eq{phi-1}--\Eq{phi-3} is that~$\Phi(h,l,q,r)$  mimics the behavior of the product~$\tilde{A}(h,q)\tilde{A}(l,r)$: if either~$\tilde{A}(h,q)$ or~$\tilde{A}(l,r)$ are close to~$0$, then \Eq{phi-1} and \Eq{phi-2} guarantee that $\Phi(h,l,q,r)$~will also be close to zero; if both values are close to one, then \Eq{phi-3} allows also  $\Phi(h,l,q,r)$~to be close to one. 

Another product between variables, i.e., a term in the form~$A(h,q)\mu(q)$, appears in \Eq{am}. Following a similar approach, we introduce a set of new variables, $\psi(h,q)$, mimicking the ratio between the~$A(h,q)\mu(q)$ product and the host capacity~$\kappa_h$. We then  impose:
\begin{equation}
\label{eq:psi-1}
\psi(h,q)\leq \tilde{A}(h,q),\quad\forall h\in\Hc,q\in\Qc;
\end{equation}
\begin{equation}
\label{eq:psi-2}
\sum_{q\in\Qc}\psi(h,q)\leq 1,\quad\forall h\in\Hc \,,
\end{equation}
which mimic \Eq{am}.  
By replacing all products between $\tilde{A}$-variables with a $\Phi$-variable and all products between~$\tilde{A}$- and $\mu$-variables with a $\psi$-variable, we obtain a {\em convex problem}, which can efficiently be solved through commercial solvers.

\subsubsection{Steps 3--4: Z-score and decisions}

Let us assume that no VNF has been placed yet. We then solve an instance of the convex problem described in \Sec{maxz-convex}, and use the values of the variables~$\tilde{A}(h,q)$ and~$\psi(h,q)$ to decide which VNF to place at which host.

Recall that $\tilde{A}(h,q)$~is the relaxed version of our placement variable~$A(h,q)$, so we would be inclined to use that to make our decision. However, we also need to account for how much computational capacity VNFs would get, as expressed by~$\psi(h,q)$. If such a value falls below~the threshold $T_\psi(h,q)=\frac{\Lambda(q)}{\kappa_h}$, then VNF~$q$ may not be able to process the incoming requests, i.e., constraint \Eq{mu-gt-lambda} may be violated.

To prevent this, we define our Z-score, i.e., how confident we are about placing VNF~$q$ at host~$h$, as follows:
\begin{equation}
\label{eq:Z}
Z(h,q)=\tilde{A}(h,q)+\ind_{\left [\psi(h,q)\geq T_\psi(h,q)\right ]} \,,
\end{equation}
where $\ind$ is the indicator function. 
Recalling that $\tilde{A}$-values are constrained between~$0$ and~$1$, favoring high values of  \Eq{Z} means that we prefer a deployment that results in $\psi$-values greater than the threshold, if such a deployment exists. Otherwise, we make the placement decision based on the $\tilde{A}$-values only.

Specifically, we select the host~$h^\star$ and VNF~$q^\star$ associated with the maximum Z, i.e.,~$h^\star,q^\star\gets\arg\max_{h\in\Hc,q\in\Qc}Z(h,q)$, and place VNF~$q^\star$ in host~$h^\star$. We fix this decision and repeat the procedure till all VNFs are placed (i.e., we perform  exactly $|\Qc|$~iterations).

We now present two example runs of MaxZ, for two scenarios with different inter-host latencies.

\noindent\rule{1\columnwidth}{0.5mm}
\begin{example}
    \label{ex:maxz-lowdelay}
    Consider a simple case with two hosts~$\Hc=\{h_1,h_2\}$ with the same  CPU capacity~$\kappa_h=\SI{5}{requests/s}$,  two VNFs~$\Qc=\{q_1,q_2\}$, and only one request class~$k$ with~$\lambda_k=\SI{1}{requests/\second}$. Requests need to subsequently traverse~$q_1$ and~$q_2$. The inter-host latency~$\delta(h_1,h_2)$ is set to~$\SI{5}{\milli\second}$, while $D^\text{QoS}=\SI{50}{\milli\second}$. Then, intuitively,  the optimal solution is to deploy one VNF per host.

We  solve the problem in \Sec{maxz-convex}.
After the first iteration, we obtain~$\mathbf{\tilde{A}}=\left [\begin{smallmatrix} 0.5 & 0.5\\ 0.5 & 0.5 \end{smallmatrix}\right ]$, $\mathbf{\psi}=\left [\begin{smallmatrix} 0.5 & 0.5\\ 0.5 & 0.5 \end{smallmatrix}\right ]$, and $\mathbf{Z}=\left [\begin{smallmatrix} 1.5 & 1.5\\ 1.5 & 1.5 \end{smallmatrix}\right ]$
    \footnote{In all matrices, rows correspond to hosts and columns to VNFs.}. In such a case, using a tie-breaking rule, we place VNF~$q_1$ at host~$h_1$. 
    In the second iteration, we have $\mathbf{\tilde{A}}=\left [\begin{smallmatrix} 1 & 0.38\\ 0 & 0.62 \end{smallmatrix}\right ]$, $\mathbf{\psi}=\left [\begin{smallmatrix} 0.8 & 0.19\\ 0 & 0.61 \end{smallmatrix}\right ]$, and $\mathbf{Z}=\left [\begin{smallmatrix} 2 & 1.38\\ 0 & 1.62 \end{smallmatrix}\right ]$. We ignore the entries pertaining to VNF~$q_1$ that has been already placed and, since~$Z(h_2,q_2)>Z(h_1,q_2)$, we deploy VNF~$q_2$ at host~$h_2$, which corresponds to the intuition that, given the small value of $\delta$, VNFs should be spread across the hosts.
\end{example}
\noindent\rule{1\columnwidth}{0.5mm}
\noindent\rule{1\columnwidth}{0.5mm}
\begin{example}
    \label{ex:maxz-highdelay}
    Let us now consider the same scenario as in \Ex{maxz-lowdelay}, but assume a much longer latency~$\delta(h_1,h_2)=\SI{100}{\milli\second}$. The best solution will now be to place both VNFs at the same host.

    After the first iteration, we obtain~$\mathbf{\tilde{A}}=\left [\begin{smallmatrix} 0.7 & 0.7\\ 0.3 & 0.3 \end{smallmatrix}\right ]$, $\mathbf{\psi}=\left [\begin{smallmatrix} 0.5 & 0.5\\ 0.3 & 0.3 \end{smallmatrix}\right ]$, and $\mathbf{Z}=\left [\begin{smallmatrix} 1.7 & 1.7\\ 1.3 & 1.3 \end{smallmatrix}\right ]$. Again using a tie-breaking rule, we place VNF~$q_1$ at host~$h_1$.
    In the second iteration, we have $\mathbf{\tilde{A}}=\left [\begin{smallmatrix} 1 & 0.7\\ 0 & 0.2 \end{smallmatrix}\right ]$, $\mathbf{\psi}=\left [\begin{smallmatrix} 0.6 & 0.4\\ 0 & 0.2 \end{smallmatrix}\right ]$, and $\mathbf{Z}=\left [\begin{smallmatrix} 2 & 1.8\\ 0 & 1.2 \end{smallmatrix}\right ]$. We again ignore the entries in the first column and, since~$Z(h_1,q_2)>Z(h_2,q_2)$, we place VNF~$q_2$ at host~$h_1$, making optimal decisions.
\end{example}
\noindent\rule{1\columnwidth}{0.5mm}

\subsection{CPU allocation}
\label{sec:sol-mu}

Once the MaxZ heuristic introduced in \Sec{sol-maxz} provides us with deployment decisions, we need to decide the CPU allocation, i.e., the values of the $\mu(q)$~variables in the original problem described in \Sec{problem}. This can be achieved simply by solving the problem in \Eq{obj} 
but keeping the deployment decision fixed, i.e., replacing the $A(h,q)$~variables with parameters whose values come from the MaxZ heuristic. Indeed, we can prove the following property. 
\begin{property}
\label{prop:convex}
If the deployment decisions are fixed, then the problem of optimizing \Eq{obj} subject to 
\Eq{am}--\Eq{bandwidth} is convex.
\end{property}
\begin{IEEEproof}
Several constraints of the original problem only involve $A(h,q)$~variables, and thus simply become conditions on the input parameters: this is the case of \Eq{onehost}, \Eq{rates}, and \Eq{bandwidth}. Also, constraints \Eq{am} and \Eq{mu-gt-lambda} are linear in the variables~$\mu(q)$. With regard to the objective function,  \Eq{dk1} is now linear with respect to $\mu(q)$, while  \Eq{resptime} is convex, even if it does not look so {\em prima facie}. Indeed, its second derivative is~$\frac{\dv}{\dv^2\mu(q)}\frac{1}{\mu(q)-\lambda(q)}=\frac{2}{\left(\mu(q)-\Lambda(q)\right)^3}$, which is positive for any~$\mu(q)>\Lambda(q)$. That condition is required for system stability; therefore, we can conclude that constraint \Eq{resptime} is convex over the all region of interest.
Finally, the objective function in \Eq{obj} is in min-max form, which preserves convexity.
\end{IEEEproof}

\Prop{convex} guarantees that we can make our CPU allocation decisions, i.e., decide on the $\mu(q)$ values, in polynomial time. We can further enhance the solution efficiency by reducing the optimization problem to the resolution of a system of equations, through the Karush-Kuhn-Tucker (KKT) conditions.

\subsubsection{KKT conditions}
\label{sec:kkt}

In several nonlinear problems, including convex ones, optimal solutions are guaranteed to have certain properties, known as KKT conditions~\cite{kkt}. This greatly simplifies and speeds up the search for the optimum, as such a search can be restricted to solutions satisfying the KKT conditions.

The KKT conditions are:
\begin{enumerate}
    \item stationarity;
    \item primal feasibility;
    \item dual feasibility;
    \item complementary slackness.
\end{enumerate}
Stating them requires associating (i) re-writing the objective and constraints in {\em normal form}, and (ii) associating a {\em KKT multiplier} with each of the constraints.

Therefore, we introduce
an auxiliary variable~$\rho$ representing the maximum $\frac{D_k}{D_k^\text{QoS}}$ ratio, and imposing that for each service class~$k\in\Kc$:
\begin{eqnarray}
\label{eq:honor-r}
\rho  \hspace{-2mm}& \mathord{\geq} &  \hspace{-2mm}\frac{D_k}{D_k^{\text{QoS}}}\nonumber \\
 \hspace{-2mm}& \mathord{=} &  \hspace{-2mm}\frac{1}{D_k^{\text{QoS}}} \left(
\sum_{q\in\Qc}\frac{\gamma_k(q)}{\mu(q)-\Lambda(q)}+ \right . \nonumber \\
 \hspace{-2mm}&  &  \hspace{-4mm}\left. \sum_{q,r\in\Qc}  \hspace{-2mm}\gamma_k(q)\PP(r|q,k)  \hspace{-3mm}\sum_{h_1,h_2\in\Hc} \hspace{-3mm} A(h_1,q)A(h_2,r)\delta(h_1,h_2)\right) \,.
\end{eqnarray}
At this point, the objective is simply to minimize~$\rho$.

We also need to re-write constraints \Eq{am}, \Eq{mu-gt-lambda} and \Eq{honor-r} in normal form, and associate to them the multipliers~$M_q$, $M_h$ and~$M_k$ respectively. The resulting Lagrangian function is:
\begin{equation}
\label{eq:lagrangian}
L=\rho+\sum_{q\in\Qc}M_qX_q+\sum_{h\in\Hc}M_hY_h+\sum_{k\in\Kc}M_kW_k,
\end{equation}
where:
\begin{equation}
\label{eq:x}
X_q=-\mu(q)+\Lambda(q);
\nonumber \end{equation} \begin{equation}
Y_h= \sum_{q\in\Qc}A(h,q)\mu(q)-\kappa_h; \nonumber
\end{equation}

\begin{multline}
W_k=\sum_{q\in\Qc}\frac{\gamma_k(q)}{D_k^{\text{QoS}}}\frac{1}{\mu(q)-\Lambda(q)}+\\
\sum_{q,r\in\Qc}  \hspace{-2mm}\gamma_k(q)\PP(r|q,k)  \hspace{-3mm}\sum_{h_1,h_2\in\Hc} \hspace{-3mm} A(h_1,q)A(h_2,r)\frac{\delta(h_1,h_2)}{D_k^{\text{QoS}}}-\rho. \nonumber
\end{multline}

Stationarity, the first KKT condition, requires that~$\nabla_{r\rho\mu(q)}L=0$,
which translates into the following equations:
\begin{align}
\label{eq:kkt1-k}
\frac{\partial}{\partial \rho}L=0\iff 1-\sum_{k\in\Kc}M_k=0.
\end{align}

Furthermore, for each~$q\in\Qc$, we have:
\begin{align}
\label{eq:kkt1-mu}
\frac{\partial}{\partial \mu(q)}L=0\iff
-M_q+\sum_{h\in\Hc}M_hA(h,q)+\\
\nonumber
-\sum_{k\in\Kc}\frac{M_k\gamma_k(q)}{D_k^{\text{QoS}}}\frac{1}{\left(\mu(q)-\Lambda(q)\right)^2}
=0
\end{align}

The primal feasibility condition requires that all constraints are met. The third one, dual feasibility, is only relevant for problems that contain equality constraints, which is not our case.

Finally, complementary slackness requires that either the inequality constraints are active, or the corresponding multipliers are zero, i.e., 
\begin{align}
\label{eq:slackness-q}
M_qX_q=0,\quad\forall q\in\Qc,\\
\label{eq:slackness-h}
M_hY_h=0,\forall h\in\Hc,\\
\label{eq:slackness-k}
M_kW_k=0,
\quad\forall k\in\Kc \,.
\end{align}

Based on \Eq{slackness-q}--\Eq{slackness-k}, the multipliers assume the following meaning:
\begin{itemize}
    \item $M_q$ is zero for all {\em stable} queues, i.e., the queues fulfilling the condition~$\mu(q)>\Lambda(q)$;
    \item $M_h$ is zero for all {\em non-strained} hosts, i.e., hosts used for less than their CPU capacity~$\kappa_h$;
    \item $M_k$ is zero for all {\em non-critical} classes, i.e., classes for which the $\frac{D_k}{D_k^{\text{QoS}}}$ ratio is strictly lower than~$\rho$.
\end{itemize}

We can now determine the {\em global} computational complexity of our approach, including the VNF placement through the MaxZ heuristic and the CPU allocation by optimizing \Eq{obj}. 
\begin{property}
\label{prop:complexity}
Our solution strategy, including the MaxZ VNF placement heuristic in \Sec{sol-maxz} and the CPU allocation strategy in \Sec{sol-mu} has polynomial computational complexity.
\end{property}
\begin{IEEEproof}
Running the MaxZ heuristic requires solving exactly~$|\Qc|$ convex problems, and the complexity of doing so is cubic in the number of variables, which in turn is linear in~$|\Hc|$ and~$|\Qc|$.
It follows that the total complexity of MaxZ is~$O\left(|\Qc|(|\Qc||\Hc|)^3\right)=O(\max\left\{|\Qc|^4,|\Hc|^3\right\})$.
MaxZ also dominates the total computational complexity, because deciding the CPU allocation as described in \Sec{sol-mu}, requires only solving a system of equations, which has~\cite{barrett1994templates} cubic complexity in the number~$|\Qc|$ of variables.
\end{IEEEproof}

\section{Special case: full-load conditions}
\label{sec:fully}

In this section, we seek to further reduce the complexity of the CPU allocation problem.
%
Let us start from the Lagrange multipliers derived earlier, and recall that 
we require {\em stability}, i.e.,~$\Lambda(q) < \mu(q)$, hence~$M_q=0$ for all queues~$q\in\Qc$. 
  
Given the above and \Eq{kkt1-mu}, we can write that, for each queue~$q\in\Qc$ deployed at host $h$,
\begin{equation}
\label{eq:ktt1-simpler}
M_h=\sum_{k\in\Kc}M_k\frac{\gamma_k(q)}{D_k^{\max}}\frac{1}{\left(\mu(q)-\Lambda(q)\right)^2}.
\end{equation}

Recalling the meaning of the multipliers, we can state the following lemma and properties.
\begin{lemma}
\label{lemma:obvious}
If CPU assignment decisions are made optimizing the objective \Eq{obj}, then there exists at least one critical class, i.e., for which equality holds in \Eq{honor-r}.
\end{lemma}
\begin{IEEEproof}
Constraint \Eq{honor-r} must be active for at least one~$k\in\Kc$, otherwise, the selected value of~$r$ would not be optimal.
\end{IEEEproof}

\begin{property}
\label{prop:k-ergo}
All hosts traversed by service requests of critical classes are strained.
\end{property}
\begin{IEEEproof}
Let~$k$ be a critical class (hence,~$M_k>0$), $q$ be a host serving it, and~$H(q)$ the host~$q$ is deployed at. From \Eq{ktt1-simpler}, it follows:
\begin{equation}
\nonumber
M_{H(q)}\geq M_k\frac{\gamma_k(q)}{D_k^{\max}}\frac{1}{\left(\mu(q)-\Lambda(q)\right)^2}.
\end{equation}
The quantity at the second member is positive ($M_k>0$ because class~$k$ is strained, and~$\gamma_k(q)>0$ because by hypothesis clients of class~$k$ are served at~$q$). This implies that~$M_{H(q)}>0$, and therefore, by \Eq{slackness-h}, that host~$h$ is strained.
\end{IEEEproof}

\begin{figure}
\centering
\psfrag{l1}[c]{$\lambda_2$}
\psfrag{l2}[c]{$\lambda_1$}
\psfrag{q1}{$q$}
\psfrag{c1}{$h$}
\includegraphics[width=.2\textwidth]{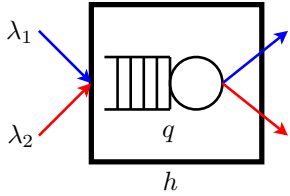}
\caption{
A simple system where two classes of clients traverse the same queue. Host~$h$ will always be strained; additionally, depending on the values of~$D_k^{\max}$, either one or both the classes will be critical.
\label{fig:small}
} 
\end{figure}
\begin{figure}
\centering
\includegraphics[width=.35\textwidth]{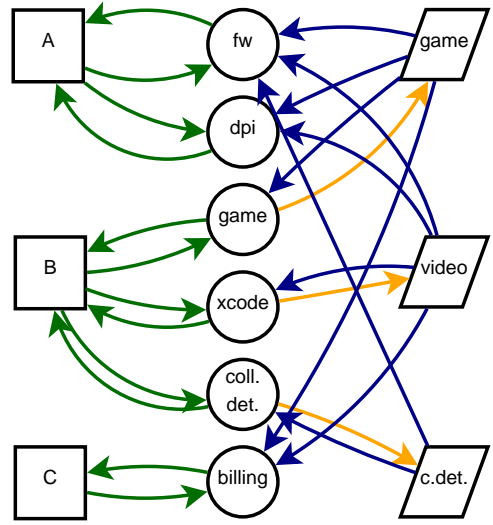}
\caption{
The graph~$\Gc$ generated for the system depicted in \Fig{example}. Left, center and right edges correspond to hosts, VNFs and classes respectively. Green edges are created according to rule (ii), blue edges according to rule (iii), and yellow edges according to rule (iv).
\label{fig:example-graph}
} 
\end{figure}

\begin{property}
\label{prop:h-ergo}
VNFs deployed at a strained host serve at least one critical class each.
\end{property}
\begin{IEEEproof}
Let us consider a queue~$q$, By hypothesis, its host~$H(q)$ is strained, i.e.,~$M_{H(q)}>0$. From \Eq{ktt1-simpler}, it follows:
\begin{equation}
\nonumber
\sum_{k\in\Kc}M_k\frac{\gamma_k(q)}{D_k^{\max}}\frac{1}{\left(\mu(q)-\Lambda(q)\right)^2}>0,
\end{equation}
which can only be if there is at least one class~$k$ that is critical (hence~$M_k>0$) and whose clients are served by~$q$ (hence~$\gamma_k(q)>0$).
\end{IEEEproof}

In summary,
we are guaranteed that there is
at least one critical class, and that {\em all} the hosts it traverses are strained, and that {\em each} of the VNFs deployed on the strained hosts (not only the ones serving  requests of the original critical class) serve at least one critical class. This can lead to a cascade effect, as shown in \Ex{cascade}.

\noindent\rule{1\columnwidth}{0.5mm}
\begin{example}
\label{ex:cascade}
Consider the case in \Fig{example}. By \Lem{obvious}, at least one class is critical; let us assume that such a class is collision detection. From \Prop{k-ergo}, all hosts traversed by collision detection requests, i.e., hosts~$h$ and~$l$, are strained. Since host~$l$ is strained, from \Prop{h-ergo} it follows that each of its VNFs serves at least one critical class. Since the transcoder queue only serves the video class, the video class is critical. Similarly, since the game server only serves the game class, that class is critical as well. Finally, both video and game classes traverse host~$m$; therefore, by \Prop{k-ergo}, that host is critical as well.
\end{example}
\noindent\rule{1\columnwidth}{0.5mm}

The cascade effect shown in \Ex{cascade} might lead us to conjecture that all classes are critical and all hosts are strained. However, this is not true in general. A simple counterexample is represented in \Fig{small}, where two classes share the same queue. By \Lem{obvious}, one of the two classes will be critical, and, hence, by \Prop{k-ergo}, host~$h$ will be strained. \Prop{h-ergo} tells us what we already know, i.e., that one of the two classes will be critical, but it  does not imply that {\em both} will be. Indeed, that depends on the values of~$D_k^{\max}$: if~$D_{k_1}^{\max}=D_{k_2}^{\max}$, then both classes are critical; otherwise, the class with the lowest~$D_k^{\max}$ value will be critical and the other will not. 

However, we can state a {\em sufficient} condition for all classes to be critical (and, hence, all hosts to be strained), {\em regardless} the ~$D_k^{\max}$ values. It is based on (i) building a graph~$\Gc$ representing the hosts, VNFs and classes in our system (as shown in \Fig{example-graph}), and (ii) verifying a simple property over it.
\begin{theorem}
\label{thm:sufficient}
Let~$\Gc=(\Vc,\Ec)$ be a directed graph where:\\
(i) there is a vertex for every host, queue, and class, i.e.,~$\Vc=\Hc\cup\Qc\cup\Kc$;\\
(ii) for every host~$h$ and queue~$q$ s.t.~$\mathbf{A}(h,q)=1$, add to~$\Ec$ a pair of edges~$(q,h)$ and~$(h,q)$;\\
(iii) for every queue~$q$ and class~$k$ s.t.~$\gamma_k(h)>0$, add to~$\Ec$ an edge~$(k,q)$;\\
(iv) for every queue~$q$ and class~$k$ s.t.~$\gamma_k(h)>0$ and~$k$ is the only class using~$q$, i.e.,~$\gamma_j=0,\forall j\neq k$, add to~$\Ec$ an edge~$(q,k)$.\\
If graph~$\Gc$ is strongly connected, then all classes in~$\Kc$ are critical and all hosts in~$\Hc$ are strained.
\end{theorem}
\begin{IEEEproof}
\Lem{obvious} guarantees us that there is at least one critical class~$k^\star$; let us then start walking through the graph from the corresponding edge and mark all vertices we can reach as critical (if corresponding to classes) or strained (if corresponding to hosts). 
Through edges added according to rule (ii) and (iii), we will be able to reach all hosts traversed by clients of the critical class, and those hosts will be strained as per \Prop{k-ergo}. 
Edges outgoing from the host vertices, created according to rule (ii), will make us reach all queues deployed at these hosts. By \Prop{h-ergo}, each of these queues serves at least one critical class. If this class is unique, i.e., if we have an edge created according to rule (iv), then those classes are critical as well, and we can repeat the process.

The strong connectivity property implies that we can reach all vertices (including all classes and all hosts) from any vertex of~$\Gc$, including the one critical class whose existence is guaranteed by \Lem{obvious}.
\end{IEEEproof}

\begin{figure}
\centering
\psfrag{h1}{$h$}
\psfrag{h2}{$h$}
\psfrag{q1}{$q$}
\psfrag{q2}{$q$}
\psfrag{k1}{$k_1$}
\psfrag{k2}{$k_2$}
\includegraphics[width=.3\textwidth]{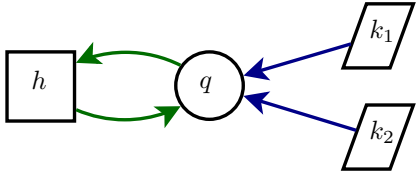}
\caption{
The graph~$\Gc$ generated for the system in \Fig{small}, which is not strongly connected (it is impossible to reach~$k_1$ from~$k_2$).
\label{fig:counter}
} 
\end{figure}

\Fig{example-graph} presents the graph~$\Gc$ resulting from the system in \Fig{example}, which is strongly connected. \Fig{counter} presents the graph for the system in \Fig{small}, which is not strongly connected and, thus, it does not meet the sufficient condition stated in \Thm{sufficient}. Recall that, because that condition is sufficient but not necessary,~$k_1$ and~$k_2$ could still be both critical, depending on their $D_k^{\max}$~values.

In scenarios like the one in \Ex{cascade}, where all classes are critical and all hosts are strained, we have:
\begin{multline}
\label{eq:busy-k}
\sum_{q\in\Qc}\frac{\gamma_k(q)}{D_k^{\text{QoS}}}\frac{1}{\mu(q)-\Lambda(q)} +\\
+\sum_{h_1,h_2\in\Hc}\sum_{q,r\in\Qc}A(h_1,q)A(h_2,r)\gamma_k(q)\PP(r|q,k)\frac{\delta(h_1,h_2)}{D_k^{\text{QoS}}}=\rho\\
 \nonumber
\quad\quad \quad\forall k\in\Kc,
\end{multline}
\begin{equation}
\label{eq:busy-h}
\sum_{q\in\Qc}A(h,q)\mu(q)=\kappa_h,\quad\forall h\in\Hc. \nonumber 
\end{equation}
The above equations can be combined with the KKT conditions stated in \Sec{kkt}, thus forcing $Y_h=0$ $\forall h\in\Hc$ and $W_k=0$ $\forall k\in\Kc$. This greatly simplifies and speeds up the process of finding the optimal CPU allocation values~$\mu(q)$. 

\section{Multiple VNF instances}
\label{sec:manyinstances}

So far, we presented our system model and solution strategy in the case where exactly one instance of each VNF has to be deployed. This is not true in general; some VNFs may need to be replicated owing to their complexity and/or load.

If the number~$N_q$ of instances of VNF~$q$ to be deployed is known, then we can replace VNF~$q$ in the VNF graph with $N_q$ replicas thereof, labeled~$q^1,q^2,\dots q^{N_q}$, each with the same incoming and outgoing edges. With regard to the $\Lambda(q)$ requests/s that have to be processed by {\em any} instance of VNF~$q$, they are split among the instances. If~$f(q,i)$ is the fraction of requests for VNF~$q$ that is processed by instance~$q^i$ (and thus~$\sum_{i=1}^{N_q}f(q,i)=1$), then instance~$q^i$ gets requests at a rate~$f(q,i)\Lambda(q)$. It is important to stress that once the $f(q,i)$~splitting fractions are known, then the resulting problem can be solved with the approach described in \Sec{solution}.

Establishing the $f(q,i)$~values is a complex problem; indeed, straightforward solutions like uniformly splitting flows (i.e.,~$f(q,i)=\frac{1}{N_q}$), are in general suboptimal. We therefore resort to a {\em pattern search}~\cite{pattern} iterative approach.

Without loss of generality, we describe our approach in the simple case~$N_q=2$. In this case, the splitting values are~$f(q,1)=f$ and~$f(q,2)=1-f$. Given an initial guess~$f_0$, an initial step~$\Delta$ and a minimum step~$\epsilon<\Delta$, we proceed as follows:
\begin{enumerate}
    \item we initialize the splitting factor~$f$ to the initial guess~$f_0$;
    \item using the procedure detailed in \Sec{solution}, we compute the objective value \Eq{obj} for the splitting values~$f$, $f+\Delta$, and~$f-\Delta$;
    \item if the best result (i.e., the lowest value of \Eq{obj}) is obtained for splitting value~$f+\Delta$ or~$f-\Delta$, then we replace~$f$ with that value and loop to step~2;
    \item otherwise, we reduce~$\Delta$ by half;
    \item if now~$\Delta$ is lower than~$\epsilon$, the algorithm terminates;
    \item otherwise, we loop to step~2.
\end{enumerate}
The intuition of the pattern-search procedure is similar, in principle, to gradient-search methods. If we find that using~$f+\Delta$ or~$f-\Delta$ instead of~$f$ produces a lower delay, then we replace the current value of~$f$ with the new one; otherwise, we try new $f$-values closer to the current one. When we are satisfied that there are no better $f$-values further  than~$\epsilon$ from the current one, the search ends.

Notice that in step~2 of our procedure we run the decision-making procedure described in \Sec{solution}; this implies that, once we find the best value of~$f$, we also know the best VNF placement and CPU allocation decisions.

From the viewpoint of complexity, the multi-instance case requires running the procedure detailed in \Sec{solution} for as many times as there are values of~$f$ to try. Since at every iteration we half the step~$\Delta$, the number of such values is at most~$\left\lceil\log_2\frac{\Delta_0}{\epsilon}\right\rceil$, $\Delta_0$~being the initial step. Such a logarithmic term does not impact the order of the global complexity, which remains polynomial as proven in \Prop{complexity}.

\section{Numerical results}
\label{sec:results}

After describing the reference topology and benchmark alternatives in \Sec{sub-ref}, this section reports the performance of MaxZ as a function of the inter-host latency and arrival rates (\Sec{sub-eff}). Next, the effects of multi-class and multi-instance scenarios (\Sec{sub-multi}) and larger-scale topologies (\Sec{sub-big}) are presented. Finally, in \Sec{sub-gen}, we generalize our results and study the number of solutions that are examined by MaxZ and its counterparts, as well as the associated running times.

\subsection{Reference scenario}
\label{sec:sub-ref}

We consider a reference topology with three hosts having CPU capacity~$\kappa_h=\SI{10}{requests/\milli\second}$ each.
As in \Ex{simple}, hosts are connected to each other through physical links
having latency that varies between \SI{50}{\milli\second} and \SI{400}{\milli\second}. For simplicity, we disregard the link capacity, i.e., we assume computation to be the bottleneck in our scenario,
as it is commonly the case in single-tenant scenarios.
Throughout our performance evaluation, we benchmark the MaxZ placement heuristic in \Sec{sol-maxz} against the following alternatives:
\begin{itemize}
    \item {\em global optimum}, found through brute-force search of all possible deployments;
    \item {\em greedy}, where the number of used hosts is minimized, i.e., VNFs are concentrated as much as possible;
    \item {\em affinity-based}~\cite{DBhamare17}, trying to place at the same host VNFs with high transition probability between them. 
\end{itemize}
After the VNF placement decisions are made, we compute the optimal CPU allocation, i.e., the optimal $\mu(q)$~values, as explained in \Sec{sol-mu}. It is important to remark that the CPU allocation strategy is the same for all placement strategies.

\begin{figure}
\centering
\includegraphics[width=1\columnwidth]{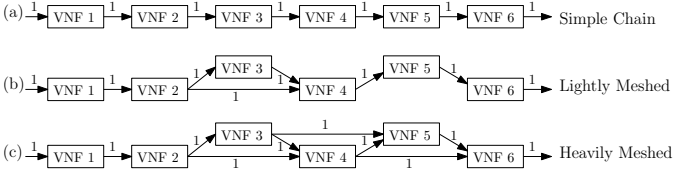}
\caption{
    The VNF graphs used in our performance evaluation, reflecting real-world service implementations.
    \label{fig:vnfgraphs}
} 
\end{figure}

\begin{figure*}
\centering
\includegraphics[width=.326\textwidth]{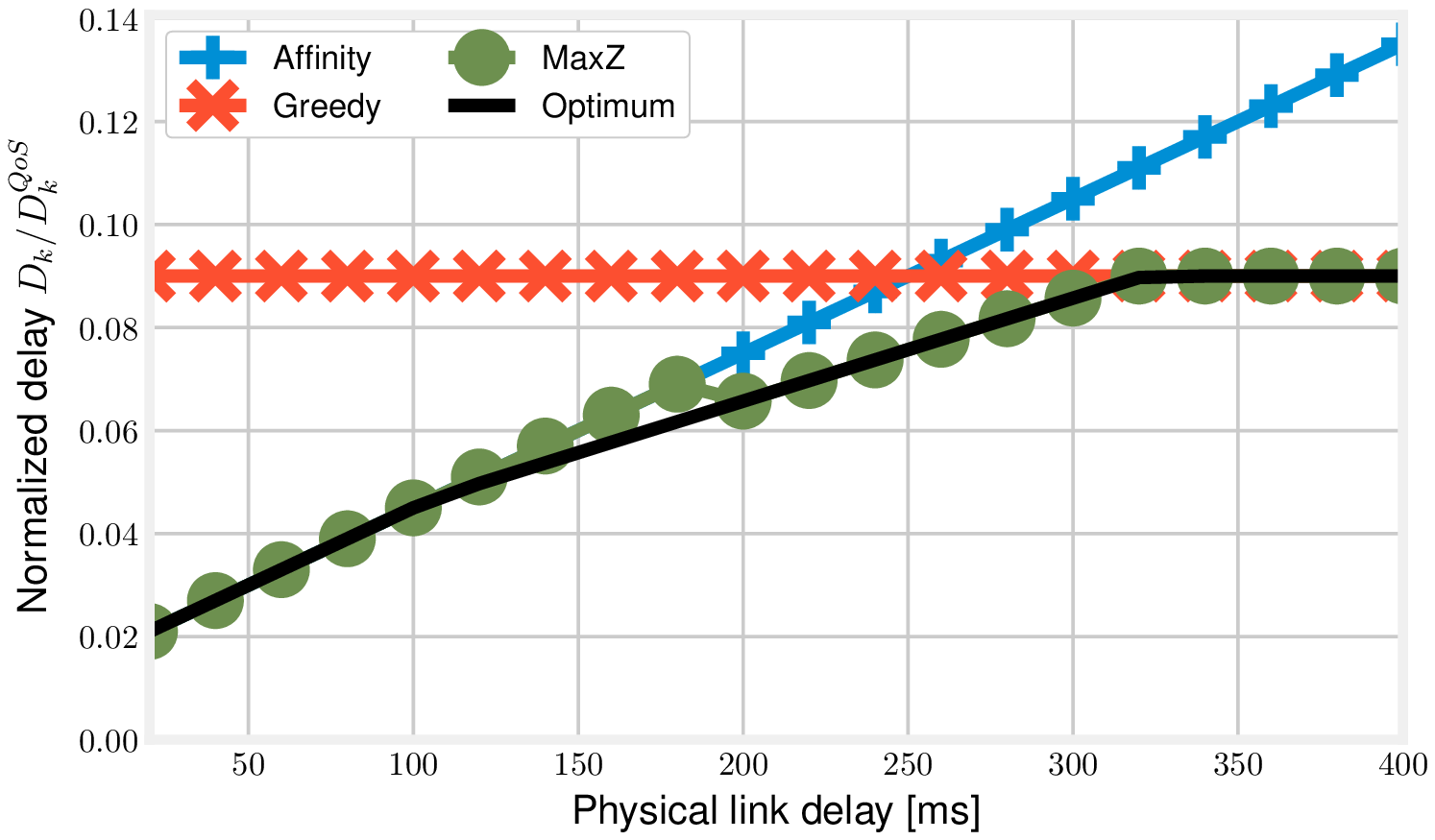}\hspace{1mm}
\includegraphics[width=.326\textwidth]{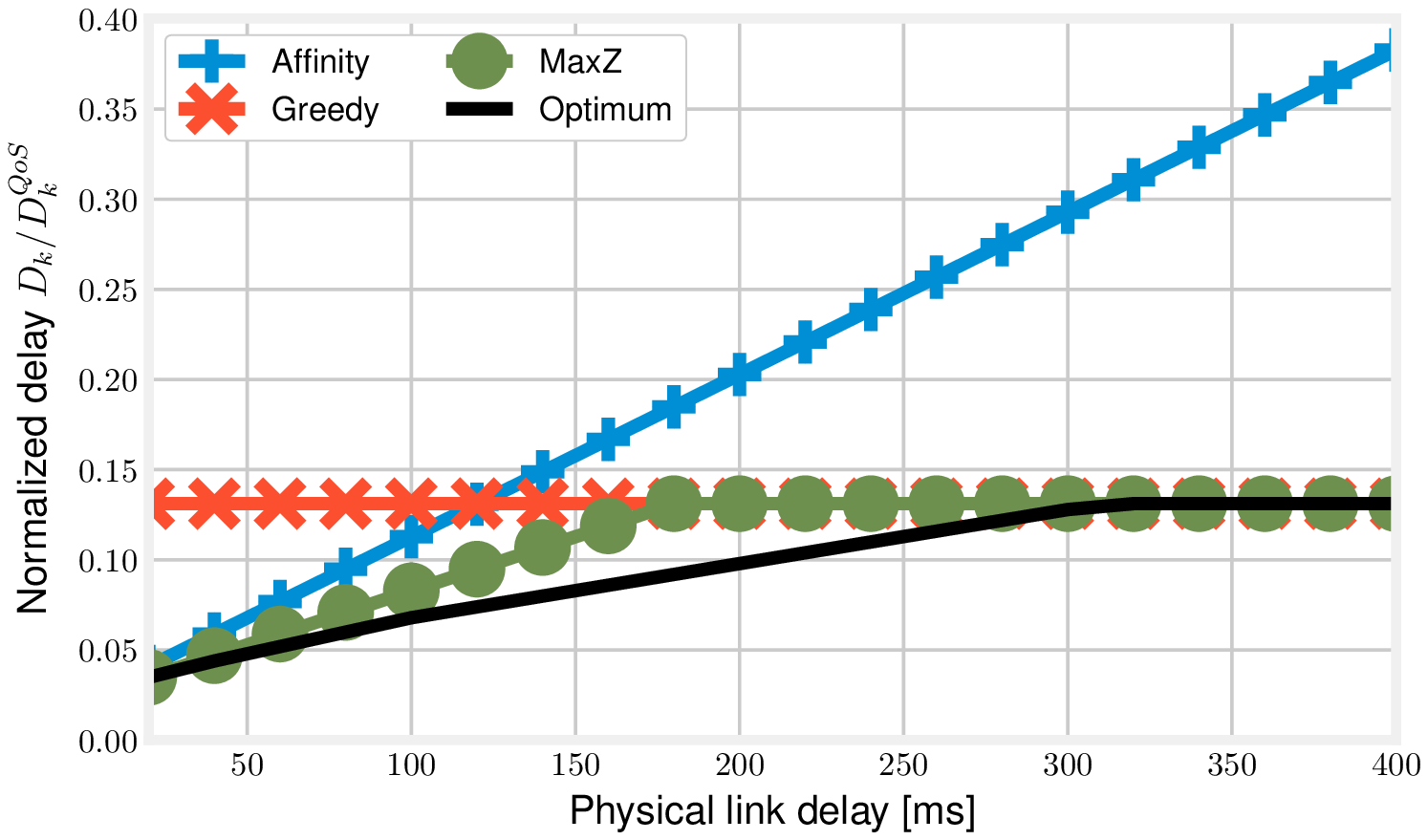}\hspace{1mm}
\includegraphics[width=.326\textwidth]{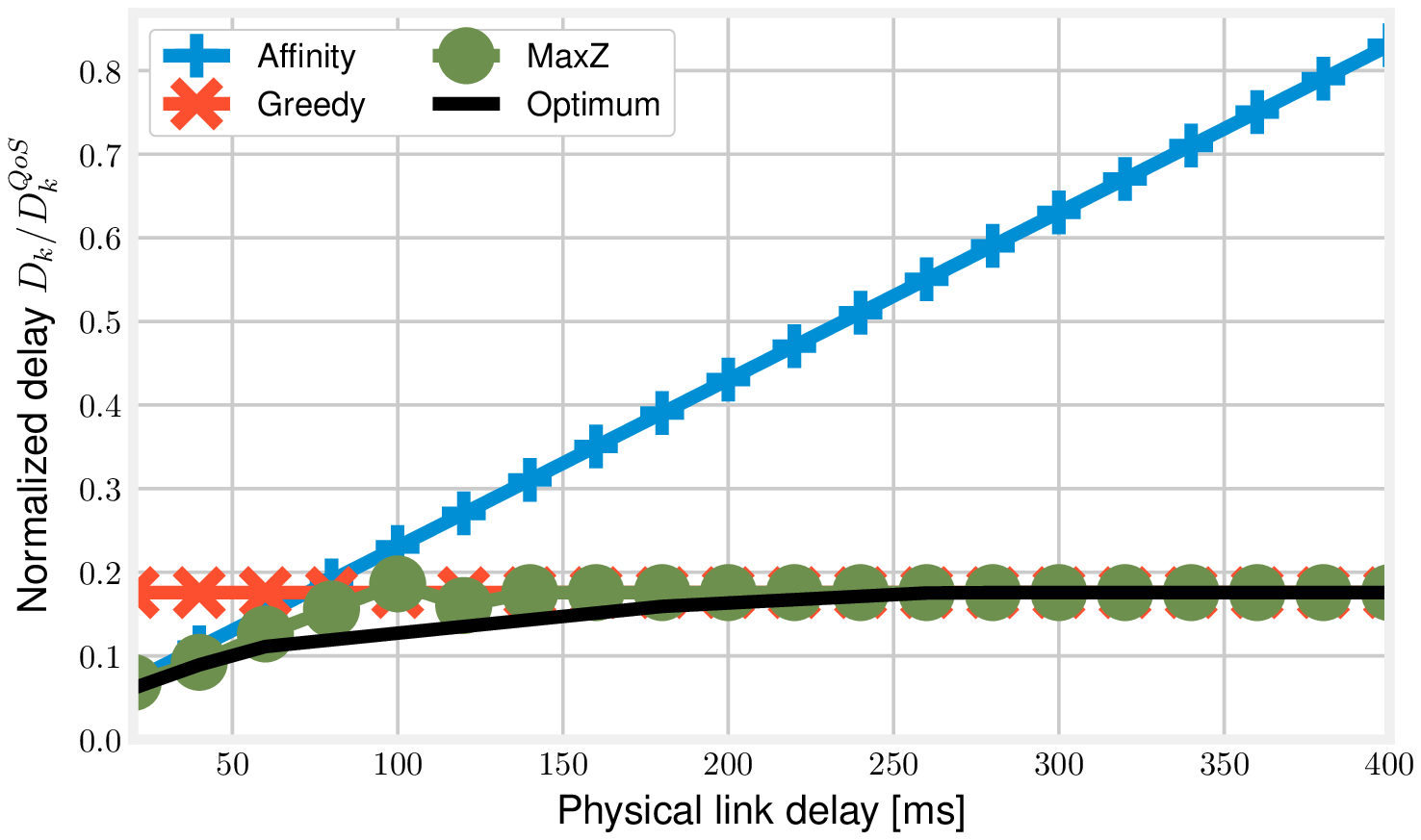}
\caption{Normalized service delay as a function of  the physical link latency, for the chain (left), light mesh (center), heavy mesh (right) VNF graphs. Note that the y-axis scale varies across the plots.   \label{fig:delta}
} 
\end{figure*}

\begin{figure}
\centering
\includegraphics[width=.9\columnwidth]{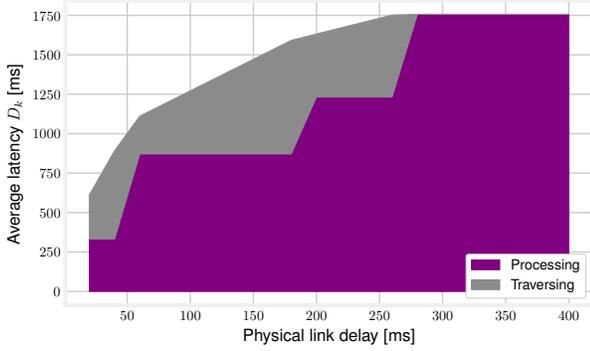}
\caption{Breakdown of the total service delay as a function of  the physical link latency, for the heavy mesh topology and the MaxZ deployment strategy.
    \label{fig:breakdown}
} 
\end{figure}

\begin{figure*}
\centering
\includegraphics[width=.326\textwidth]{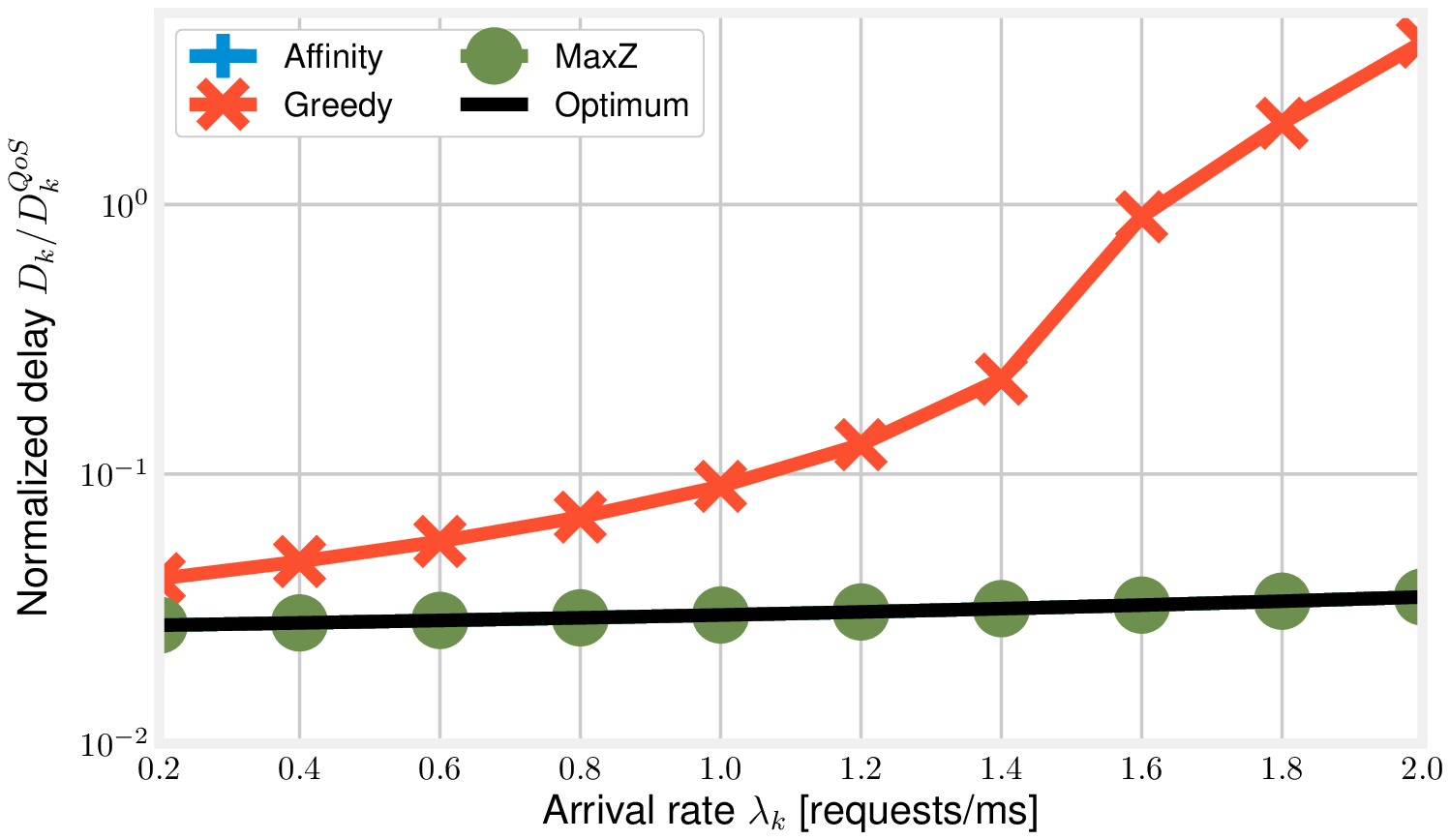}
\includegraphics[width=.326\textwidth]{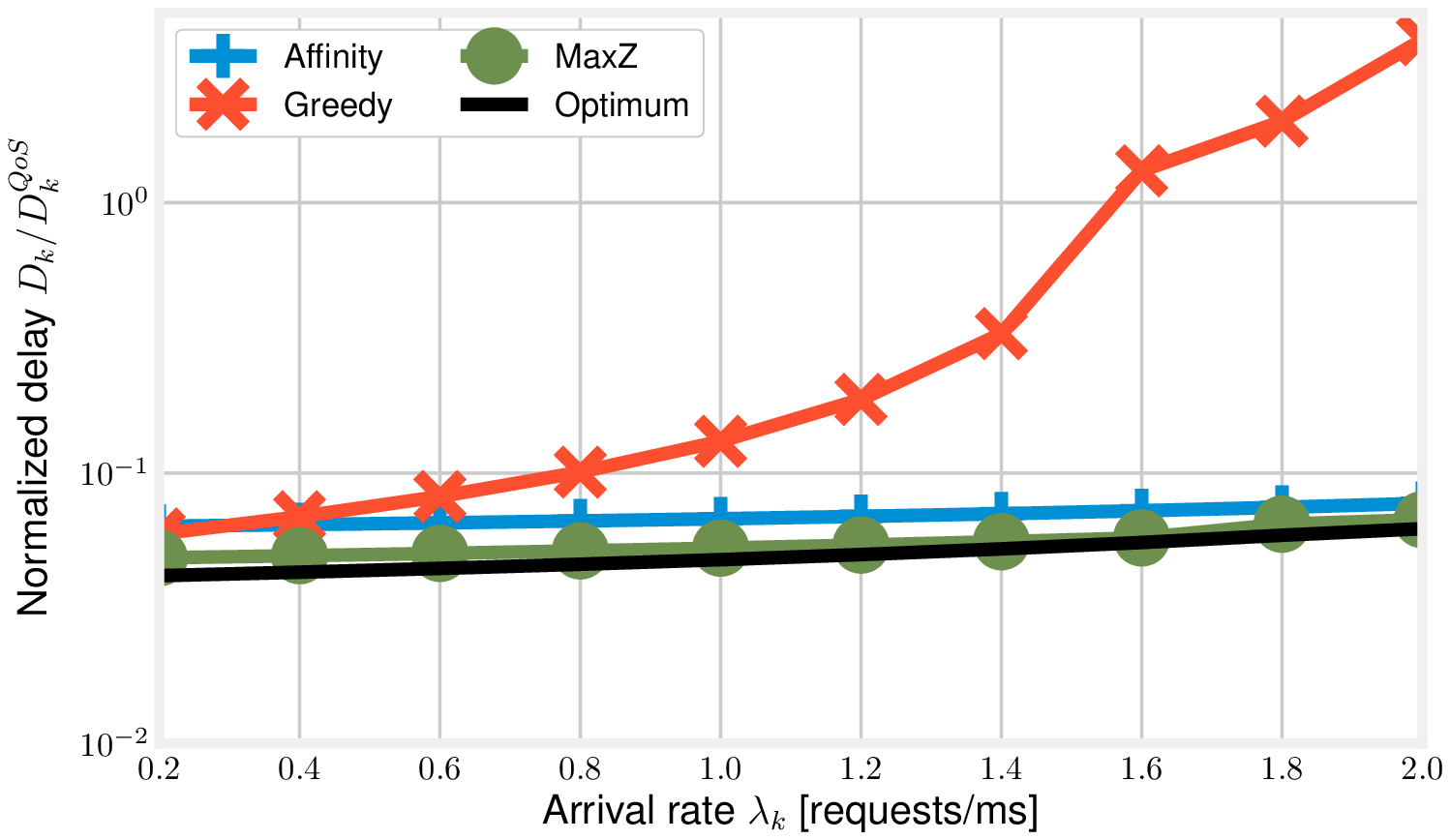}
\includegraphics[width=.326\textwidth]{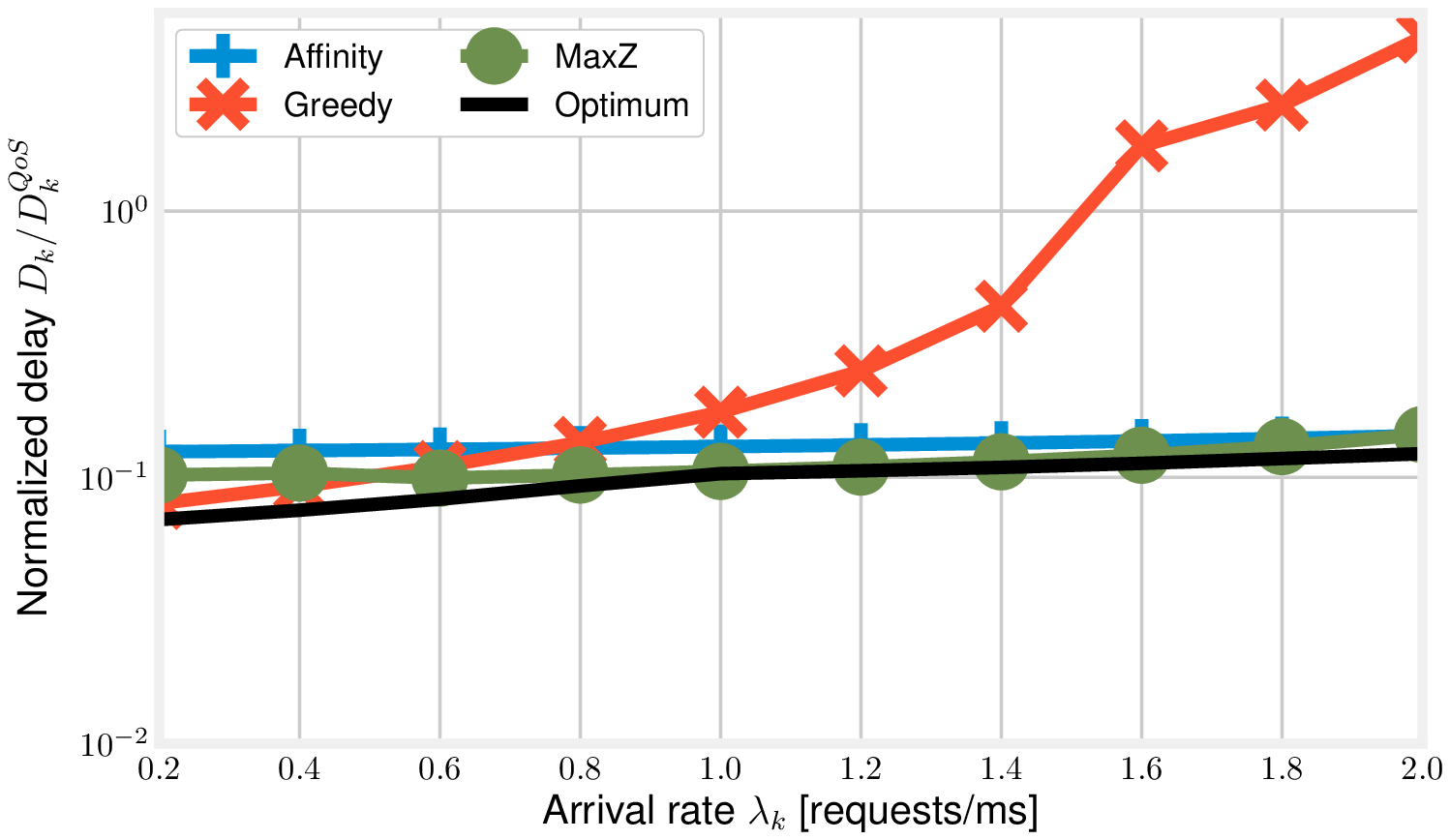}
\caption{Normalized service delay (log scale) as a function of  arrival rate~$\lambda$ for the chain (left), light mesh (center), heavy mesh (right) VNF graphs.
    \label{fig:lambda}
} 
\end{figure*}

\begin{figure*}
\centering
\includegraphics[width=.326\textwidth]{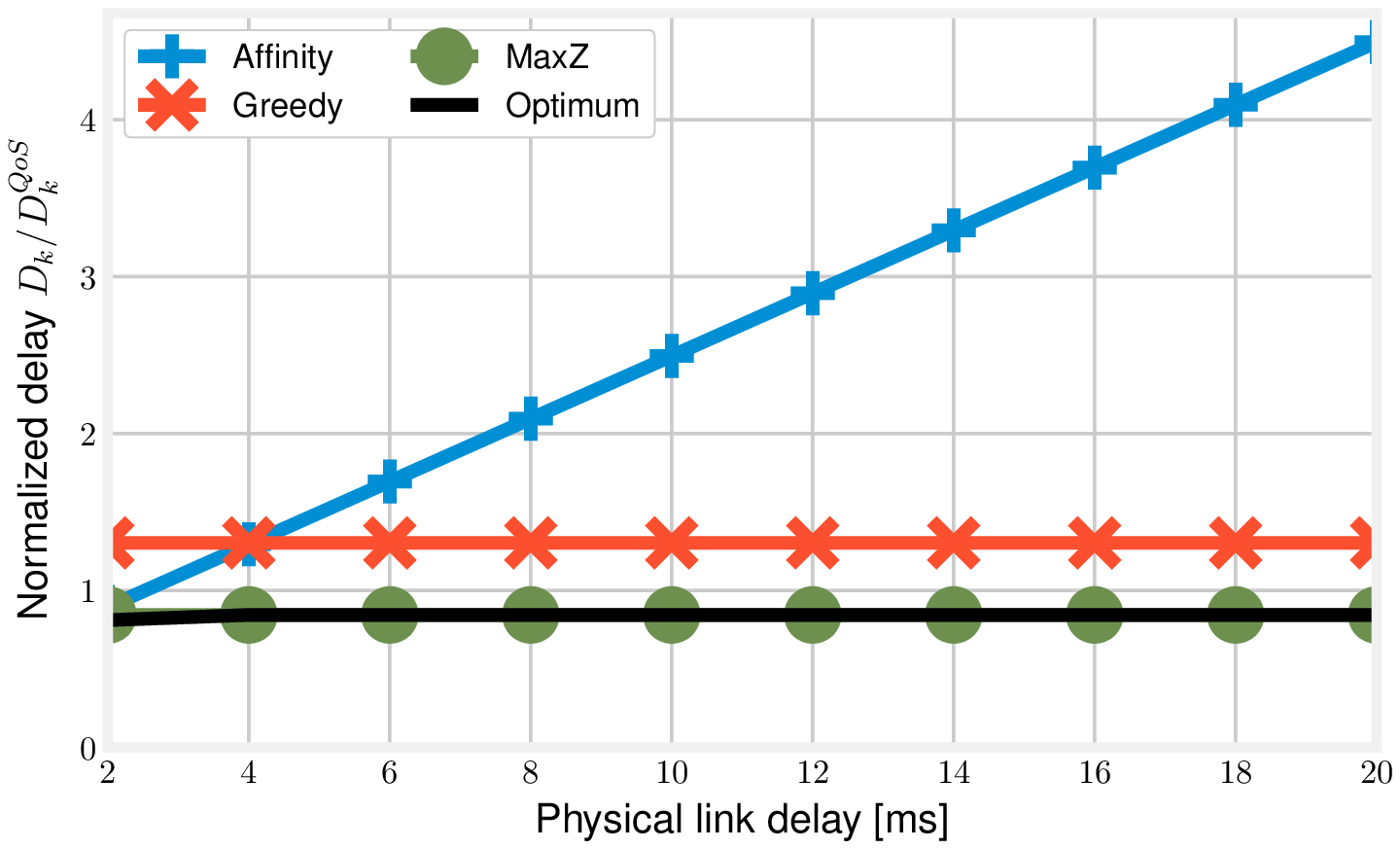}
\includegraphics[width=.326\textwidth]{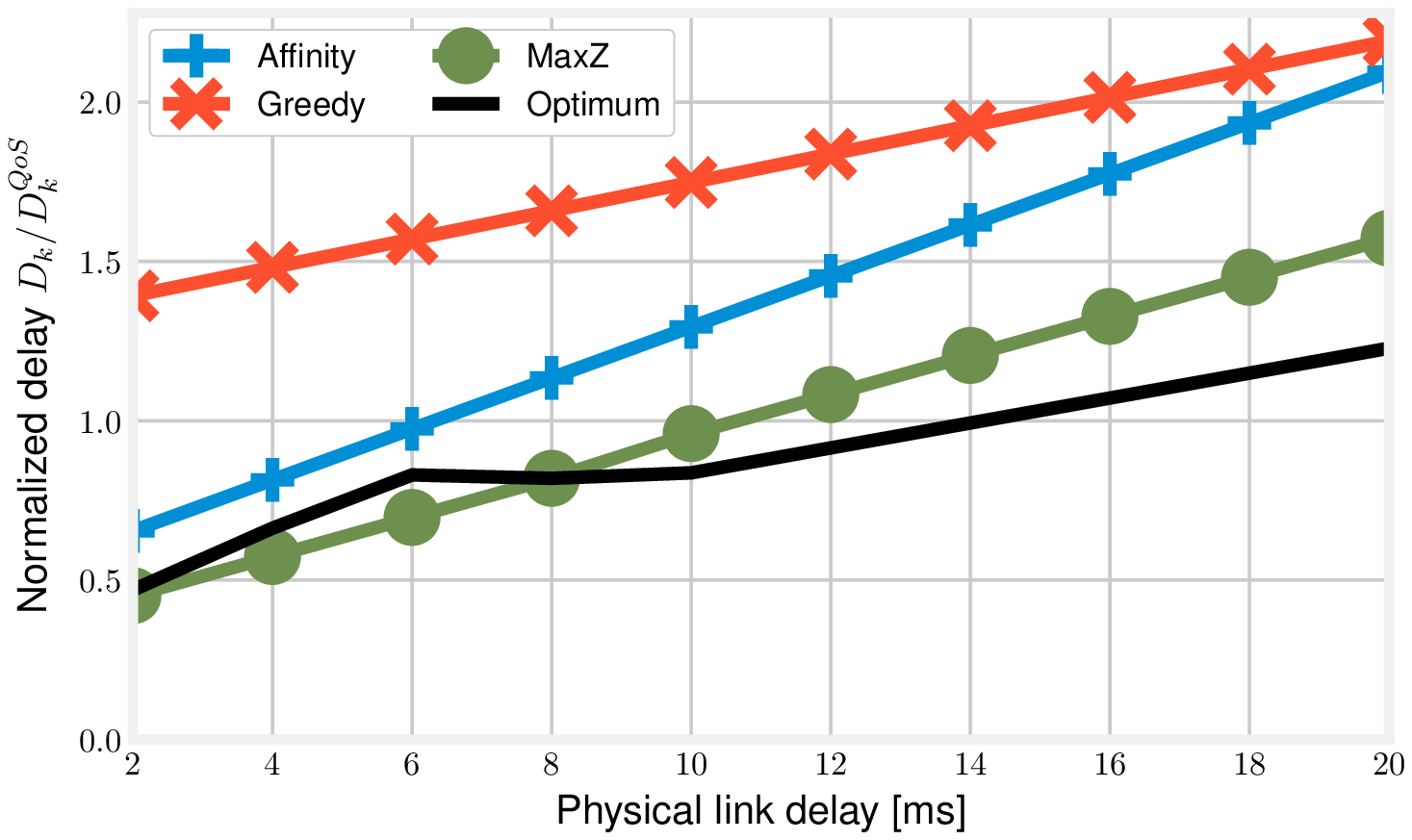}
\includegraphics[width=.326\textwidth]{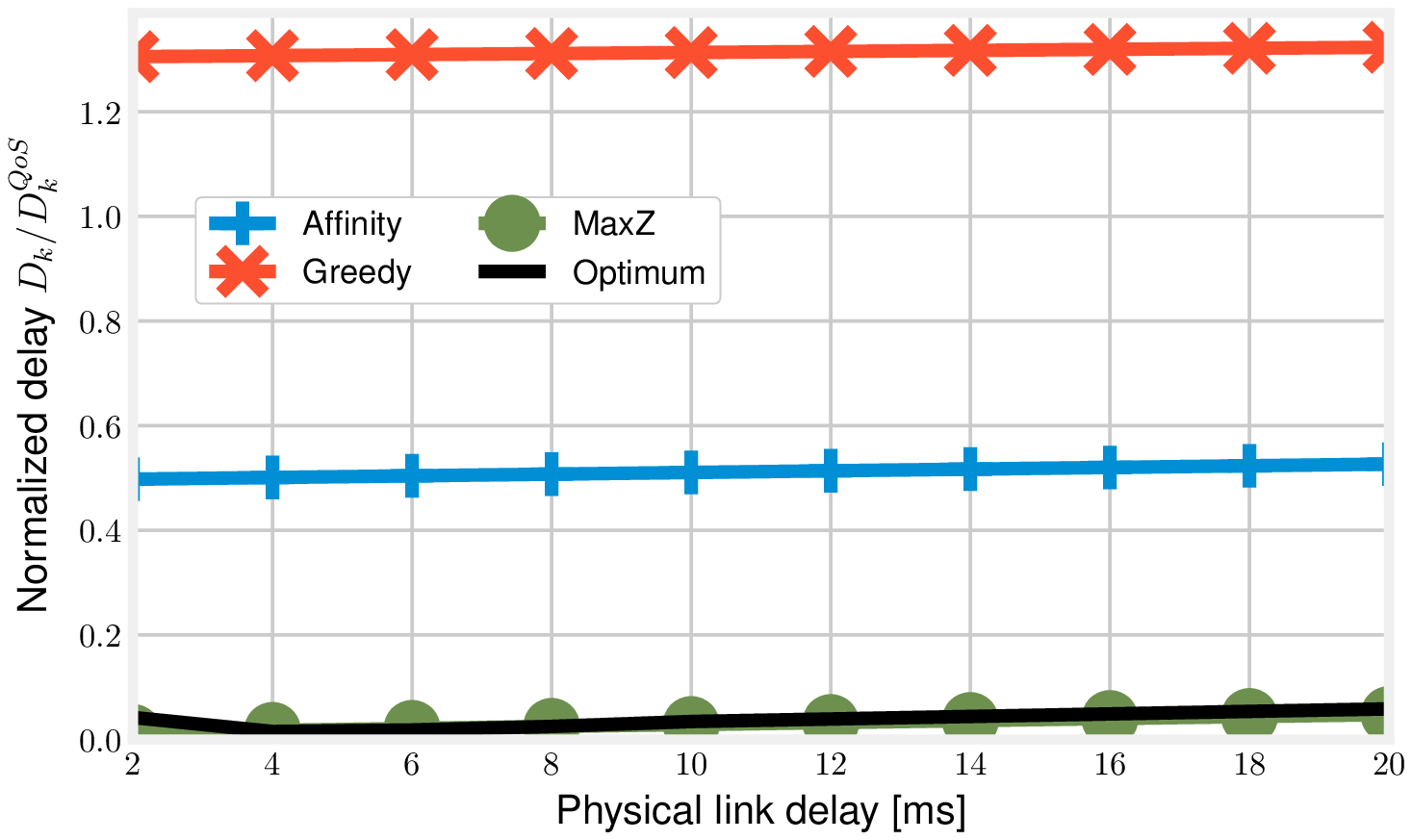}
\caption{
    Multi-class scenario, heavy mesh graph: normalized service delay vs.  arrival rate~$\lambda$ for the low-delay (left), medium-delay (center), high-delay (right) service classes.  Note that the y-axis scale varies across the plots. 
    \label{fig:multi}
} 
\end{figure*}
\begin{figure*}
\centering
\includegraphics[width=.326\textwidth]{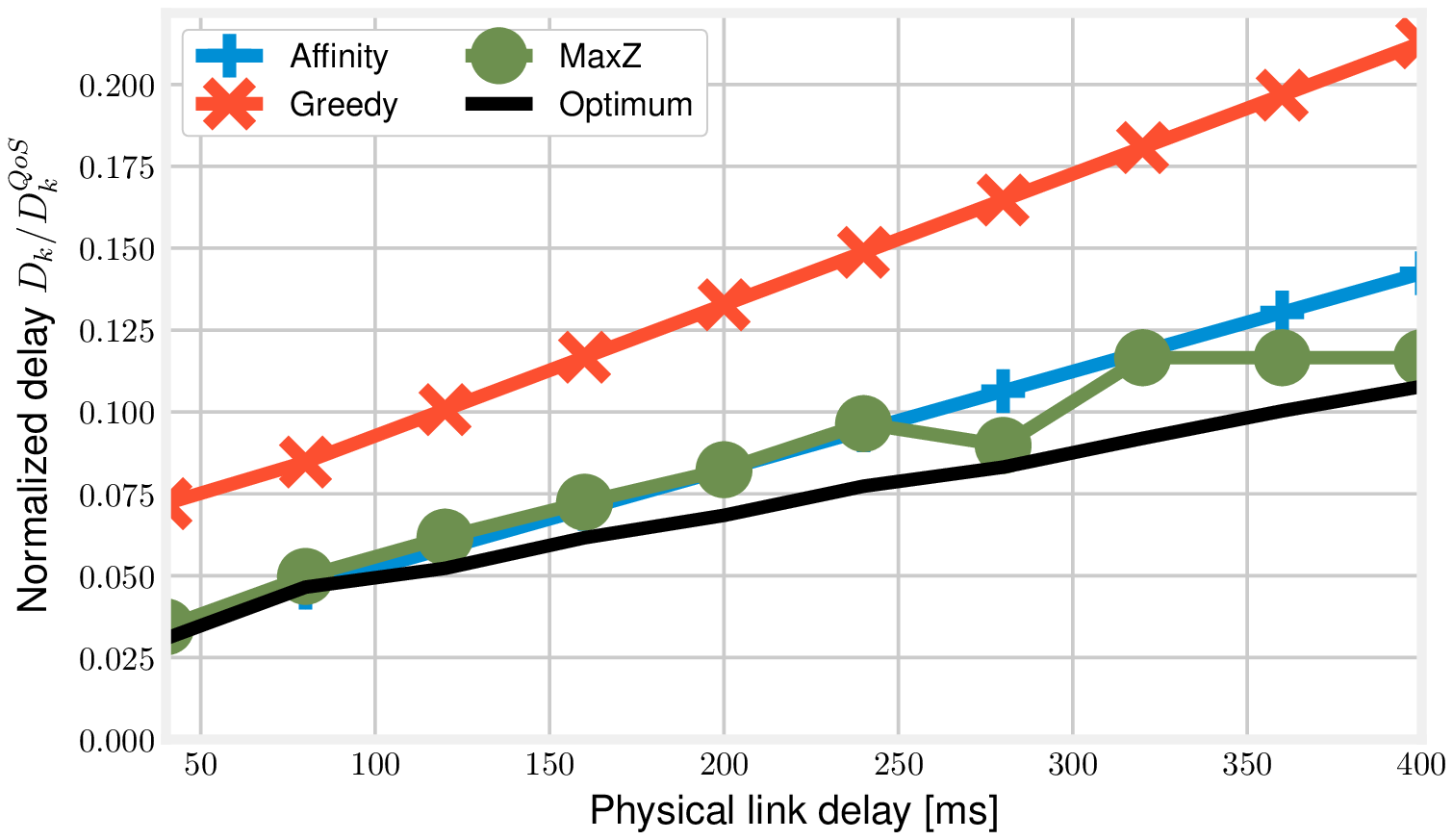}
\includegraphics[width=.326\textwidth]{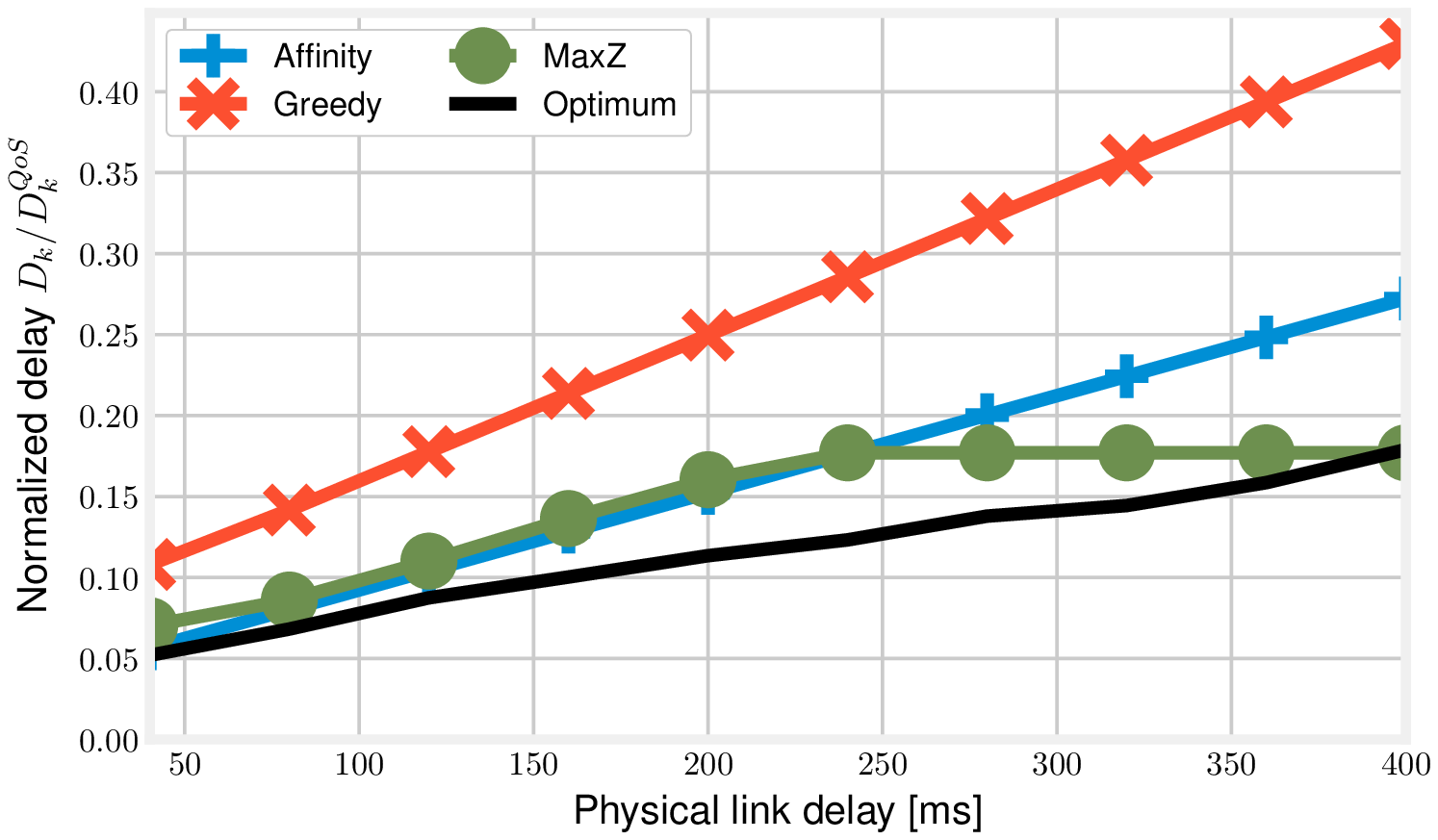}
\includegraphics[width=.326\textwidth]{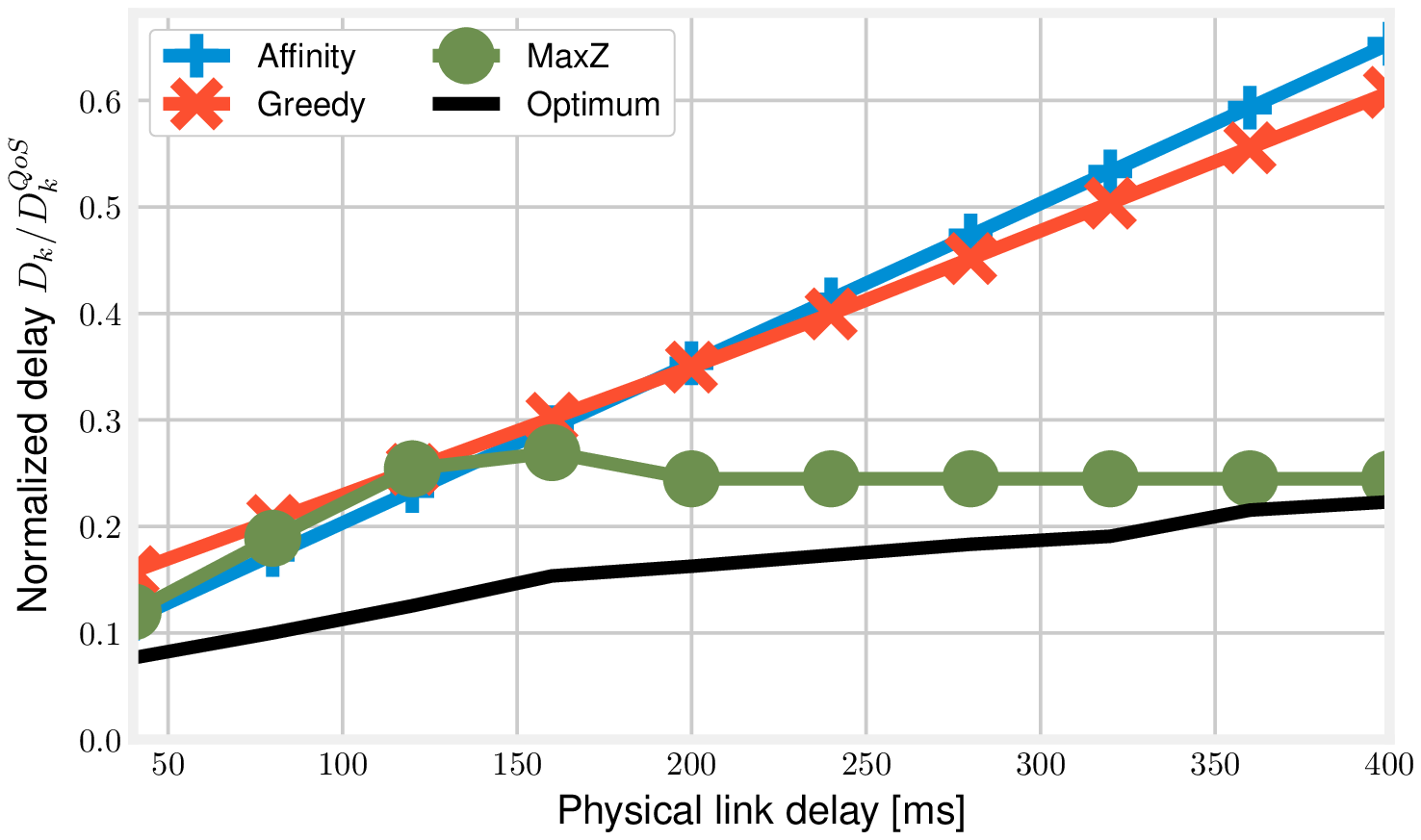}
\caption{
    Multi-instance scenario: normalized service delay vs. the physical link latency for the chain (left), light mesh (center), heavy mesh (right) VNF graphs. Note that the y-axis scale varies across the plots. 
    \label{fig:many}
} 
\end{figure*}

We first focus on a single request class~$k$, fix the arrival rate to~$\lambda_k=\SI{1}{requests/\milli\second}$, and compare the three VNF graphs depicted in \Fig{vnfgraphs}, ranging from a simple chain to a complex meshed topology. Notice how in graphs (b) and (c) requests can {\em branch} and {\em merge}, i.e., the number of requests outgoing from a VNF does not match the number of incoming ones. This is the case with several real-world functions; in particular, the light and heavy mesh topologies are akin to vEPC implementations where user- and control-plane entities are joint~\cite{tvt_mme} and split~\cite{globecom15_bearer,vepc_controller_placement},
i.e., implemented, respectively, by the same VNF or by separate ones.

\subsection{Effect of physical link latency and arrival rate}
\label{sec:sub-eff}

\Fig{delta} shows the average  service delay as a function of the physical link latency  for the VNF graphs presented in \Fig{vnfgraphs}. We can observe that the performance of Greedy is always the same regardless of such latency, as all VNFs are deployed at the same host. On the other hand, the performance of Affinity-based is quite good for low values of the physical links latency, but then quickly degrades, due to the fact that the affinity-based heuristic disregards link latency. As far as MaxZ is concerned,  it exhibits an excellent performance: it consistently yields a substantially lower service delay compared to Greedy and Affinity-based, and is always very close to the optimum.

\Fig{breakdown} focuses on the heavy mesh topology, and breaks down the service delay  yielded by MaxZ  into its computation and traversing latency components. Processing latency only depends upon the VNF placement, while traversing latency depends upon both the VNF placement (which determines how many inter-host links are traversed) and the per-link latency. When such latency is low, MaxZ tends to spread the VNFs across all available hosts, in order to assign more CPU. As the physical link latency increases, the placement becomes more and more concentrated (thus resulting in lower $\mu(q)$~values and higher processing times), until, when the link latency is very high, all VNFs are placed at the same host and there is no traversing latency at all.

\Fig{delta} and \Fig{breakdown} clearly illustrate the importance of flexible CPU allocation. If we only accounted for the minimum CPU required by VNFs, as in~\cite{infocom15_optimal,DBhamare17}, we could place all of them in the same host, as the Greedy strategy does. This would result, as we can see from the far right in \Fig{breakdown}, in high processing times and {\em two unused} hosts.

We now fix the physical link latency to~$\SI{50}{\milli\second}$, and change the arrival rate~$\lambda$ between~$\SI{0.1}{requests/\milli\second}$ and~$\SI{2}{requests/\milli\second}$; \Fig{lambda} summarizes the service delay yielded by the placement strategies we have studied. A first observation  concerns the Greedy strategy: since all VNFs are placed in the same host, as $\lambda$  increases, VNFs receive an amount of CPU that is barely above the minimum limit~$\Lambda(q)$. This, as per \Eq{resptime}, results in processing times that grow very large. The difference between the other placement strategies tends to become less significant; intuitively, this is because processing times dominate the total delay, and thus spreading the VNFs as much as possible is always a sensible solution. MaxZ still consistently outperforms Affinity-based, and performs very close to the optimum.

\begin{figure}
\centering
\includegraphics[width=1\columnwidth]{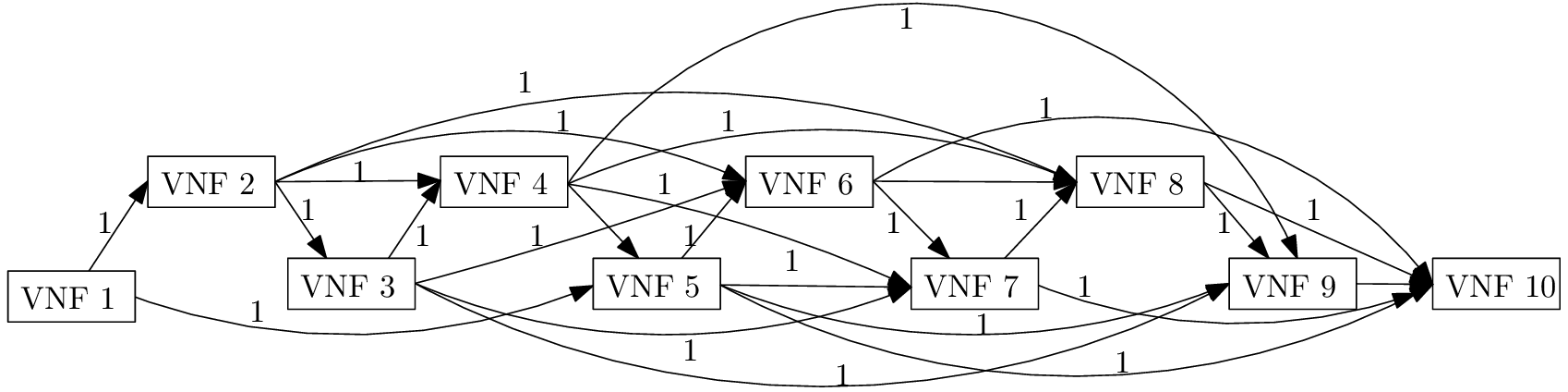}
\caption{Extreme mesh VNF graph.
\label{fig:extreme}
} 
\end{figure}

\begin{figure}
\centering
\includegraphics[width=.85\columnwidth]{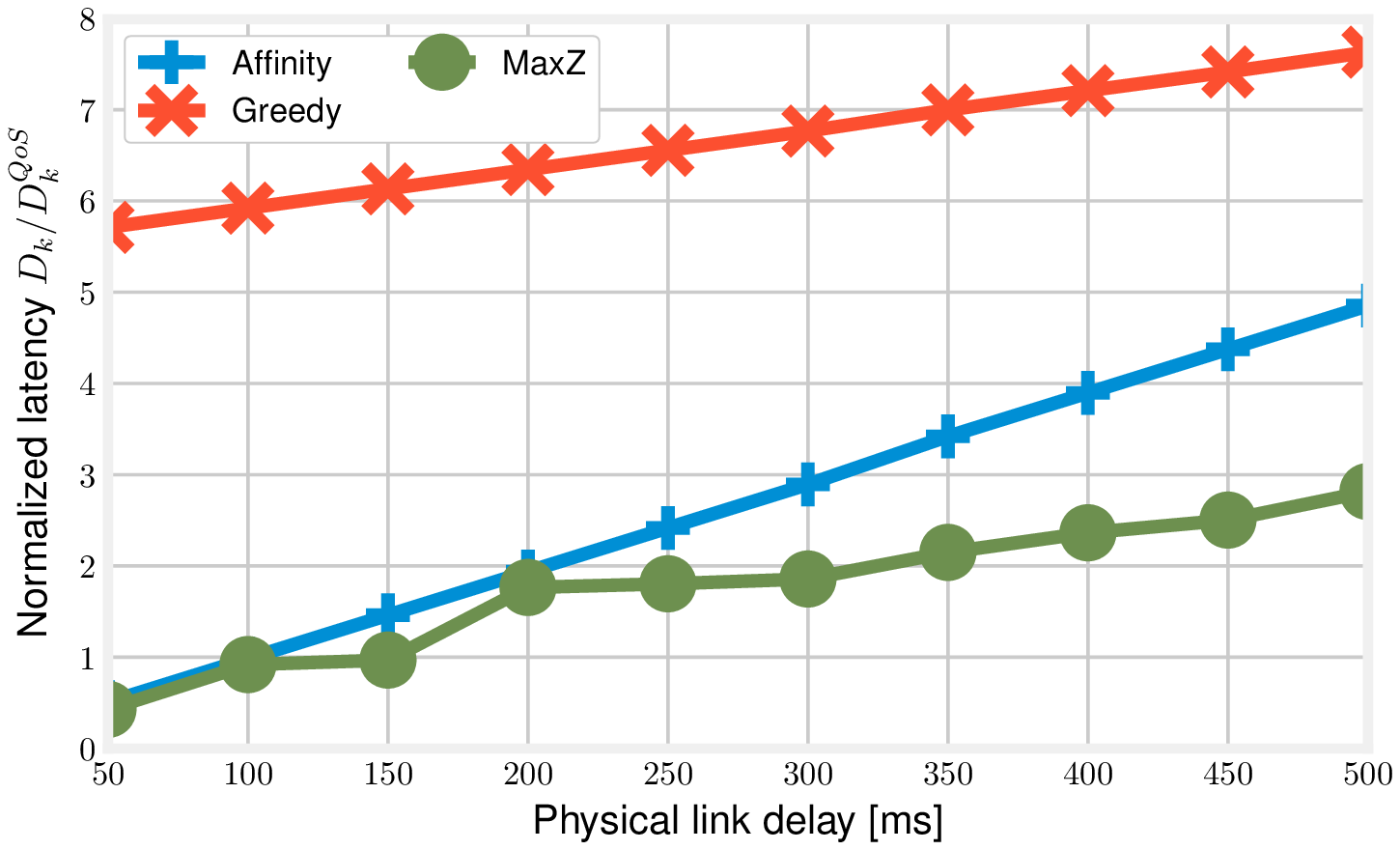}
\caption{
Extreme mesh VNF graph, 20-host topology with connectivity degree~4: normalized service delay vs. physical link latency.
\label{fig:large4}
} 
\end{figure}

\begin{figure}
\centering
\includegraphics[width=.85\columnwidth]{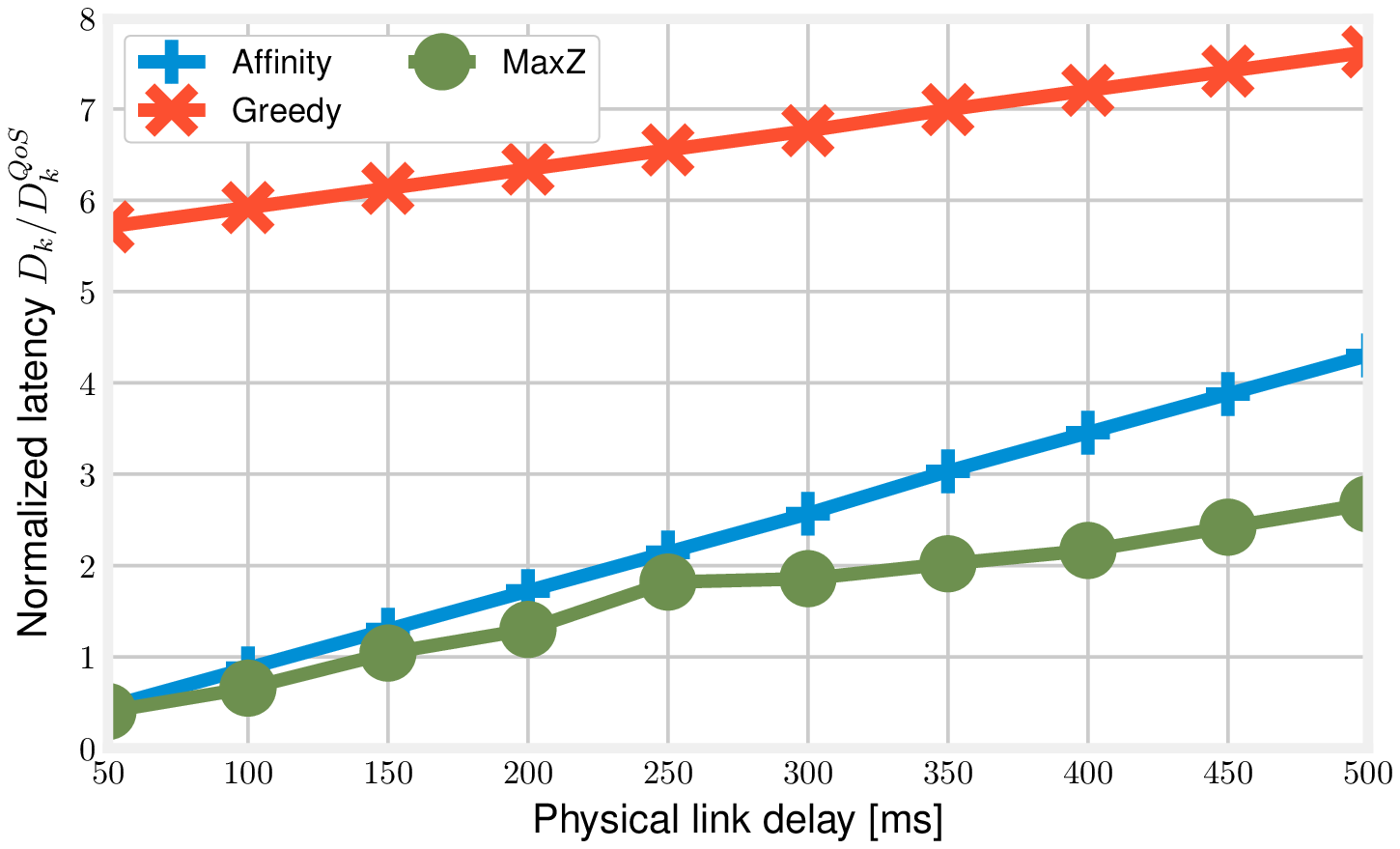}
\caption{
Extreme mesh VNF graph, 20-host topology with connectivity degree~6: normalized service delay vs. physical link latency $\delta$.
\label{fig:large6}
} 
\end{figure}

\subsection{Multiple class and VNF instances}
\label{sec:sub-multi}

In \Fig{multi}, we move to a multi-class scenario where~$|\Kc|=3$ service classes share the same VNF graph. The three classes have limit delays~$D^\text{QoS}$ of \SI{10}{\milli\second}, \SI{45}{\milli\second}, and \SI{2}{\second}, respectively corresponding to safety applications (e.g., collision detection), real-time applications (e.g., gaming), and delay-tolerant applications (e.g., video streaming). \Fig{multi} shows that all placement strategies result in delays that are roughly proportional to the limit ones. Also, the relative performance of the placement strategies remains unmodified -- MaxZ outperforms Affinity-based and is close to the optimum, while Greedy yields much higher delay. Notice that, for very high values of physical link latency, it is impossible to meet all QoS constraints, i.e., $\frac{D_k}{D_k^\text{QoS}}>1$ for at least one class~$k$. In these cases, MaxZ limits the damage by keeping the $\frac{D_k}{D_k^\text{QoS}}$ ratios as low as possible.

It is also interesting to notice in \Fig{multi}(center) that the service delay yielded by MaxZ is actually lower than the optimum. However, this does not mean that MaxZ outperforms the optimum;
indeed, due to the min-max structure of our objective \Eq{obj}, the objective value is determined by the class with the highest $\frac{D_k}{D_k^{QoS}}$ ratio. For low-to-medium link latency, such class is the low-delay one, with a normalized service delay (shown in \Fig{multi}(left)) around~$0.9$. Thus, MaxZ and the optimum strategy obtain the same objective value.

In \Fig{many}, we drop the assumption that there is only one instance of each VNF; specifically, for VNF$_4$ and VNF$_6$ we allow two instances each. By comparing \Fig{many} to \Fig{delta}, we can immediately notice that allowing multiple VNF instances substantially decreases the total delay. More interestingly, we can observe that MaxZ always outperforms its alternatives, and is very close to the optimum, except for some cases when the topology is very complex.

\subsection{Larger-scale scenarios}
\label{sec:sub-big}

We now move to larger-scale, more complex scenarios, where:
\begin{itemize}
    \item the VNF graph is the {\em extreme mesh} depicted in \Fig{extreme} and inspired to real-world VNF graphs~\cite{pimrc-wp3};
    \item there can be up to two copies of each VNF;
    \item each host is connected through physical links with either~4 or~6 other hosts chosen in such a way that  any pair of hosts is connected through one virtual link.
\end{itemize}
The fact that the physical topology is not strongly connected implies that the latency of virtual links between hosts depends on many physical links they are made of.

\Fig{large4}, obtained when each host is connected to four others, shows that the relative performance of the solution strategies does not change, with MaxZ still outperforming its alternatives. By comparing \Fig{large4} with \Fig{delta}(right), we can observe much higher service delays, often exceeding~$D_k^{\text{QoS}}$. This further highlights how MaxZ can provide near-optimal performance in all conditions, including sparsely-connected topologies where wrong placement decisions can result in very high network delays.

Moving to \Fig{large6}, obtained when each host is connected to six others, we can observe a slightly better performance for all strategies, due to shorter network delays. MaxZ is still able to clearly outperform all the alternatives, due to its ability to account for both network latencies and computation times.

\subsection{Generalization: examined solutions and running times}
\label{sec:sub-gen}

In the following, we seek to generalize the results shown above and study the number of solutions (i.e., values given to the $A(h,q)$~variables) that our benchmark approaches examine while searching for the best one. Recall that, as shown in \Prop{complexity}, MaxZ has a complexity of~$O(\max\left\{|\Qc|^4,|\Hc|^3\right\})$.

{\bf Brute-force.}
We use brute-force to seek for the globally-optimal solution. There are $|\Qc|$ number of VNFs to be placed in $|\Qc|$ hosts. Thus, in total there are $|\mathcal{H}|^{|\mathcal{Q}|}$ possible number of VNF-host placements. For each of the VNF-host placement, we obtain the optimal resource allocation for each VNF at each host by solving a convex optimization problem. Convex problems have cubic complexity in the number of variables, and each problem has~$|\mathcal{Q}|+1$ variables, corresponding one $\mu(q)$~each VNF and a variable corresponding to $r$. The problem complexity in this case is $\mathcal{O}((|\mathcal{Q}| + 1)^{3})$. Thus, the overall computational complexity of global optimization utilizing brute-force is $\mathcal{O}(|\mathcal{H}|^{|\mathcal{Q}|}(|\mathcal{Q}| + 1)^{3})$.

{\bf Greedy scheme.}
In the greedy scheme, firstly the VNFs are placed on the hosts  such that the number of hosts utilized is minimized. Thereafter, optimal resources are allocated by solving the convex optimization problem. The VNFs are first sorted in the decreasing order of their load which has the computational complexity of $\mathcal{O}(|\mathcal{Q}|\log(|\mathcal{Q}|))$. They are then allocated to $|\mathcal{H}|$ hosts, iteratively. The greedy placement algorithm has a complexity of $\mathcal{O}(|\mathcal{Q}|\log(|\mathcal{Q}|) +|\mathcal{H}|)$. As seen before, the computational complexity of resource allocation algorithm is $\mathcal{O}((|\mathcal{Q}| + 1)^{3})$. Thus, the overall computational complexity of greedy algorithm is $ \mathcal{O}(|\mathcal{Q}|\log(|\mathcal{Q}|) + |\mathcal{H}| + (|\mathcal{Q}| + 1)^{3})$, which is equivalent to $ \mathcal{O}(|\mathcal{H}| + (|\mathcal{Q}| + 1)^{3})$.

{\bf Affinity-based scheme.}
In the affinity-based scheme, firstly the VNFs are placed on the hosts such that the VNFs having high transition probability between them are placed on the same host. Thereafter, optimal resources are allocated by solving the convex optimization problem. There are $|\mathcal{Q}|$ number of VNFs to be placed in $|\mathcal{H}|$ hosts. The affinity based VNF-host placement algorithm has a complexity of $\mathcal{O}(|\mathcal{Q}||\mathcal{H}|)$. As seen before, the computational complexity of resource allocation algorithm is $\mathcal{O}((|\mathcal{Q}| + 1)^{3})$. Thus, the overall computational complexity of this scheme is $ \mathcal{O}(|\mathcal{H}||\mathcal{Q}| + (|\mathcal{Q}| + 1)^{3})$.

\begin{table}
\centering
\caption{Execution time (in seconds) for different schemes}
\label{tab:times}
\begin{tabularx}{1\columnwidth}{|X||r|r|r|r|}
\hline
 Scenario & Aff. & Greedy & MaxZ & Brute  \\
 \hline
  \hline
Base scenario (chain graph) & 0.32 & 0.34 & 4.22 & 238.83 \\ \hline
Base scenario (light mesh graph) & 0.32 & 0.29 & 4.31 & 238.26 \\ \hline
Base scenario (heavy mesh graph) & 0.35 & 0.34 & 4.53 & 246.18 \\ \hline
Base scenario (multiple classes) & 0.61 & 0.62 & 10.14 & 415.05 \\ \hline
Large scenario, 4-degree connectivity (extreme mesh graph, two instances per VNF) & 0.62 & 0.50 & 23.36 &  \\ \hline
Large scenario, 6-degree connectivity (extreme mesh graph, two instances per VNF) & 0.62 & 0.50 & 21.34 & \\ \hline
\end{tabularx}
\end{table}

{\bf Execution times.}
All the above computations refer to the order of magnitude of the worst-case computational complexity. However, it is also interesting to assess how such complexity translates into actual execution times. To this end, \Tab{times} reports the execution times of MaxZ and its counterparts for different topologies and VNF graphs, measured on a server equipped with a Xeon~E5-2600 processor and 48~GByte of RAM. We can clearly observe that, while MaxZ takes longer than the affinity-based and greedy heuristics to run, their execution times are comparable in the base scenario. Furthermore, MaxZ runs over two orders of magnitude faster than the brute-force procedure. It is also interesting to notice that the execution times in the large scenario are still limited, while the brute-force procedure is utterly unable to tackle that case.

\section{Conclusion}
\label{sec:conclusion}

We targeted the problem of orchestration in 5G networks, that requires to make decisions about VNF placement, CPU assignment, and traffic routing. We presented a queuing-based model accounting for all the main features of 5G networks, including (i) arbitrarily complex service graphs; (ii) flexible allocation of CPU power to VNFs sharing the same host, and its impact on processing time; (iii) the possibility of having multiple instances of the same VNF. 

Based on our model, we presented a methodology to make the requirement decisions jointly and effectively, based on two pillars: a VNF placement heuristics called MaxZ, and a convex formulation of the CPU allocation problem given placement decisions. We also showed, based on KKT conditions, that the CPU allocation problem is further simplified in full-load conditions, where all hosts are completely utilized.
We evaluated our methodology against multiple VNF graphs and physical topologies of varying complexity, and found the performance of MaxZ to consistently exceed that of state-of-the-art alternatives, and closely match the optimum.
Future research directions
include multi-tenant scenarios, where multiple verticals share the same infrastructure. Furthermore, future work will
aim at optimizing the number of instances to be deployed for each VNF, and designing a low-complexity heuristic for it, and will investigate other KPIs than service delay.

\section*{Acknowledgment}
This work is supported by the European Commission through the H2020 projects 5G-TRANSFORMER (Project ID 761536) and 5G-EVE (Project ID 815074).

\bibliographystyle{IEEEtran}
\bibliography{refs}

\end{document}